\documentclass[preprint2]{aastex62}

\usepackage{graphicx}
\usepackage[suffix=]{epstopdf}
\usepackage{natbib}
\usepackage{amsmath}
\usepackage{url}
\usepackage{xspace}

\newcommand{\Kepler}{\textsl{Kepler}\xspace}

\begin{document}
\title{The Evolution of Flare Activity with Stellar Age}

\shorttitle{Evolution of Flare Activity}
\shortauthors{Davenport et al.}

\correspondingauthor{James. R. A. Davenport}
\email{jrad@uw.edu}

\author{James. R. A. Davenport}
\altaffiliation{NSF Astronomy and Astrophysics Postdoctoral Fellow}
\altaffiliation{DIRAC Fellow}
\affiliation{Department of Physics \& Astronomy, Western Washington University, 516 High St., Bellingham, WA 98225, USA}
\affiliation{Department of Astronomy, University of Washington, Seattle, WA 98195, USA}

\author{Kevin R. Covey}
\affiliation{Department of Physics \& Astronomy, Western Washington University, 516 High St., Bellingham, WA 98225, USA}

\author{Riley W. Clarke}
\affiliation{Department of Physics \& Astronomy, Western Washington University, 516 High St., Bellingham, WA 98225, USA}

\author{Austin C. Boeck}
\affiliation{Department of Physics \& Astronomy, Western Washington University, 516 High St., Bellingham, WA 98225, USA}

\author{Jonathan Cornet}
\affiliation{Department of Physics \& Astronomy, Western Washington University, 516 High St., Bellingham, WA 98225, USA}

\author{Suzanne L. Hawley}
\affiliation{Department of Astronomy, University of Washington, Seattle, WA 98195, USA}

\begin{abstract}
Using a recent census of flare stars from the \Kepler survey, we have explored how flare activity evolves across stellar main sequence lifetimes. We utilize a sample of 347 stars with robust flare activity detections, and which have rotation periods measured via starspot modulations in their \Kepler light curves. 
We consider three separate methods for quantifying flare activity from optical light curves, and compare their utility for comparing flare activity between stars of differing ages and luminosities.
These metrics include: the fractional luminosity emitted in flares, the specific rate of flares emitted at a given energy, and a model for the entire flare frequency distribution. With all three approaches we find that flare activity decreases for all low-mass stars as they spin-down, and thus with age. 
Most striking is the evolution of the flare occurrence frequency distributions, which show no significant change in the power law slope with age. Since our sample is preferentially constructed of younger, more active stars, our model over-predicts the super-flare rate previously estimated for the Sun. 
Finally, we parameterize our best-fit model of the flare frequency distribution for ease in predicting the rates of flares and their associated impacts on planet habitability and detection. 
 \end{abstract}

\section{Introduction}

Magnetic activity comes in a wide range of observable phenomena, including UV and X-ray luminosity, cool starspots, and flares. Each of these signatures is observed to decline over time for low-mass stars on the main sequence. 
As the star loses angular momentum via stellar winds, the rotation velocity decreases and the internal magnetic dynamo is quieted. This age--activity connection was outlined in the seminal work by \citet{skumanich1972}, which has been directly connected with flare activity as well \citep{skumanich1986}. Magnetic activity evolution has been confirmed for low-mass stars (late F through early M type) in many studies using X-ray luminosity \citep{wright2011,nunez2017}, UV emission \citep{shkolnik2014}, Zeeman-Doppler imaging \citep{vidotto2014}, and H$\alpha$ emission \citep{west2015}, to name but a few.

This surface activity can have significant impact on the evolution of habitable zone planets orbiting active stars. Flares in particular have been studied as a possible threat to a planet's ability to retain a habitable atmosphere \citep[e.g.][]{segura2010,luger2015,tilley2017}. For example, UV flux and high energy particles impacting a terrestrial planet's atmosphere from frequent stellar flares can significantly deplete ozone, and result in a potentially uninhabitable planetary surface \citep{tilley2017}. Under more extreme scenarios, stellar activity could strip large portions of a planet's atmosphere away over timescales of a few 100 Myr \citep{luger2015}. The duration of high flare activity early in a star's life may therefore be a fundamental property in defining potential planetary habitability.

Stellar activity also makes planet searches more difficult, adding both slow variability (e.g. starspots) that can impact radial velocity studies, and fast stochastic variability (e.g. flares) that can impede transit searches \citep[e.g.][]{kipping2017}. Further, though stellar activity is less prominent at redder wavelengths, even observing in the infrared does not eliminate the potential for flares from active stars to impact transit searches, particularly for the lowest mass stars \citep{davenport2017a}. To understand planet occurrence rates for nearby low-mass stars, as well as the evolution of planetary habitability, we must constrain flare rates and properties as a function of stellar age.

Measuring the evolution of flare activity with stellar age has typically been limited to comparisons between flare stars in young clusters or stellar associations with known ages. Studying flare stars in young nearby moving groups and clusters reaches back at least five decades \citep{haro1966}, with the conclusion that all low-mass dwarf stars undergo an evolution in flare activity \citep{ambartsumian1975}. Pre-main sequence stars have also been known to exhibit high levels of flare activity compared to field-aged dwarfs \citep[e.g.][]{feigelson2001}. However, in most previous studies field stars are assumed to be approximately Solar-aged, and therefore generally flare-inactive, since few reliable age indicators for isolated field stars exist.

Space-based exoplanet transit missions such as \Kepler \citep{borucki2010} have provided a revolutionary dataset for statistical studies of flare stars, particularly for field dwarfs \citep{walkowicz2011}. Such missions are ideally suited for large scale studies of flares, as they produce high precision, continuous light curves for months to years in duration. The \Kepler survey has been used to explore ``super-flare'' activity (events with energies more than 10 times larger than those observed on the Sun) from solar-mass stars \citep{shibayama2013}, as well as from K and M dwarfs \citep{candelaresi2014}. White-light flares in \Kepler light curves have been detected across the main sequence, from massive A and F stars \citep{balona2012} down to L dwarfs \citep{gizis2013}, and produced the most detailed catalogs of flares for individual active stars to date \citep{hawley2014,davenport2014b}.

In this paper we present an ensemble analysis of flare activity in the \Kepler field, based on the flare sample amassed in \citet{davenport2016}. This sample was generated using an automated processing of the entire \Kepler light curve database, and produced a sample of over 4000 candidate flare stars. The \citet{davenport2016} catalog of flares provides the most complete census of flare activity from a large sample of field stars to date, and is the ideal dataset to study the evolution of flare rates with stellar age.

Here we explore the relationship between flare occurrence rates and stellar ages (derived from rotation periods) for stars in the \Kepler field. Our sample of flare stars with measured rotation periods is detailed in \S\ref{sec:sample}. We present three approaches for quantitatively tracing changes in flare activity over time, including modeling the flare frequency distribution as a function of both stellar mass and age in \S\ref{sec:activity}. In \S\ref{sec:model} we explore the empirical evolution of flare activity for our \Kepler sample, and provide an analytic prescription for use in other studies. Finally, a short discussion and comparison to other studies is given in \S\ref{sec:discussion}.

\section{Flare Star Sample}
\label{sec:sample}

Our sample of flare stars comes from the automated search of \Kepler light curves from \citet{davenport2016}. This study produced the most comprehensive analysis of the \Kepler field for stellar flares, processing every short (1-minute) and long (30-minute) cadence light curve in search of flares. \citet{davenport2016} produced an open-source Python flare analysis codebase named {\tt appaloosa} \citep{appaloosa_code}, designed to detrend (model) \Kepler light curves of both instrumental and astrophysical noise, detect flare candidates (positive, significant outliers), and determine the reliability of the detected flares via artificial flare injection and recovery tests. These completeness tests were run on each continuous local segment of light curve available for every star (i.e. sections of the light curve without gaps larger than 3 hours), resulting in variable completeness limits for each star as the light curve noise properties change. \citet{davenport2016} made both the {\tt appaloosa} code, and the code to generate the figures and results in the paper available online. The results presented in this paper directly extend the work of \citet{davenport2016}, and so we also make our analysis code available as an update to the {\tt appaloosa} project online.\footnote{\url{https://github.com/jradavenport/appaloosa}}

\subsection{Selecting a Robust Sample}
Since we are focused here on quantifying changes in flare activity with stellar age, we limit our analysis to stars with measured rotation periods so that stellar rotation periods can be used as a proxy for age. We note that while the rotation--age connection may be more complex than most gyrochronology prescriptions, e.g. the weakened rotational braking found in Solar-age stars from \citet{van-saders2016}, rotation nonetheless increases continuously with age (i.e. stars nominally do not speed up while on the main sequence). Rotation is therefore a good means to {\it sort} stars by their age, even if the specific age derived from gyrochronology relations is not accurate. Our work here is further insulated from these effects, as the sample of flare star candidates from \citet{davenport2016} predominantly have rotation ages of $\lesssim 1$ Gyr. We specifically adopt for our analysis the rotation period catalog of \citet{mcquillan2014}, who measured 34,040 periods from \Kepler data using an Autocorrelation Function analysis of each light curve, and did extensive testing against other period-finding approaches such as the Lomb-Scargle Periodogram.

The \citet{davenport2016} sample of $\sim$4,000 candidate flare stars contains several types of contamination from variable stars that do not exhibit flaring activity. For example, we found some eclipsing binaries and pulsating stars were able to fool the {\tt appaloosa} code, due to sharp features in their light curve, particularly at the 30-minute cadence. The worst cases of this failure appear to be caused by insufficient modeling of periodic signals by {\tt appaloosa}, namely by using too few sine curves to fit each significant period found and leaving sharp or ``peaked'' structures in the model residuals. Future versions of {\tt appaloosa} will improve on this detrending algorithm, and we urge caution when adopting the \citet{davenport2016} flare sample blindly. Our analysis is largely free from such contamination, as we require each star to have a rotation period identified by \citet{mcquillan2014}, who rejected such eclipsing and pulsating targets.

\begin{figure*}[!t]
\centering
\includegraphics[width=2.25in]{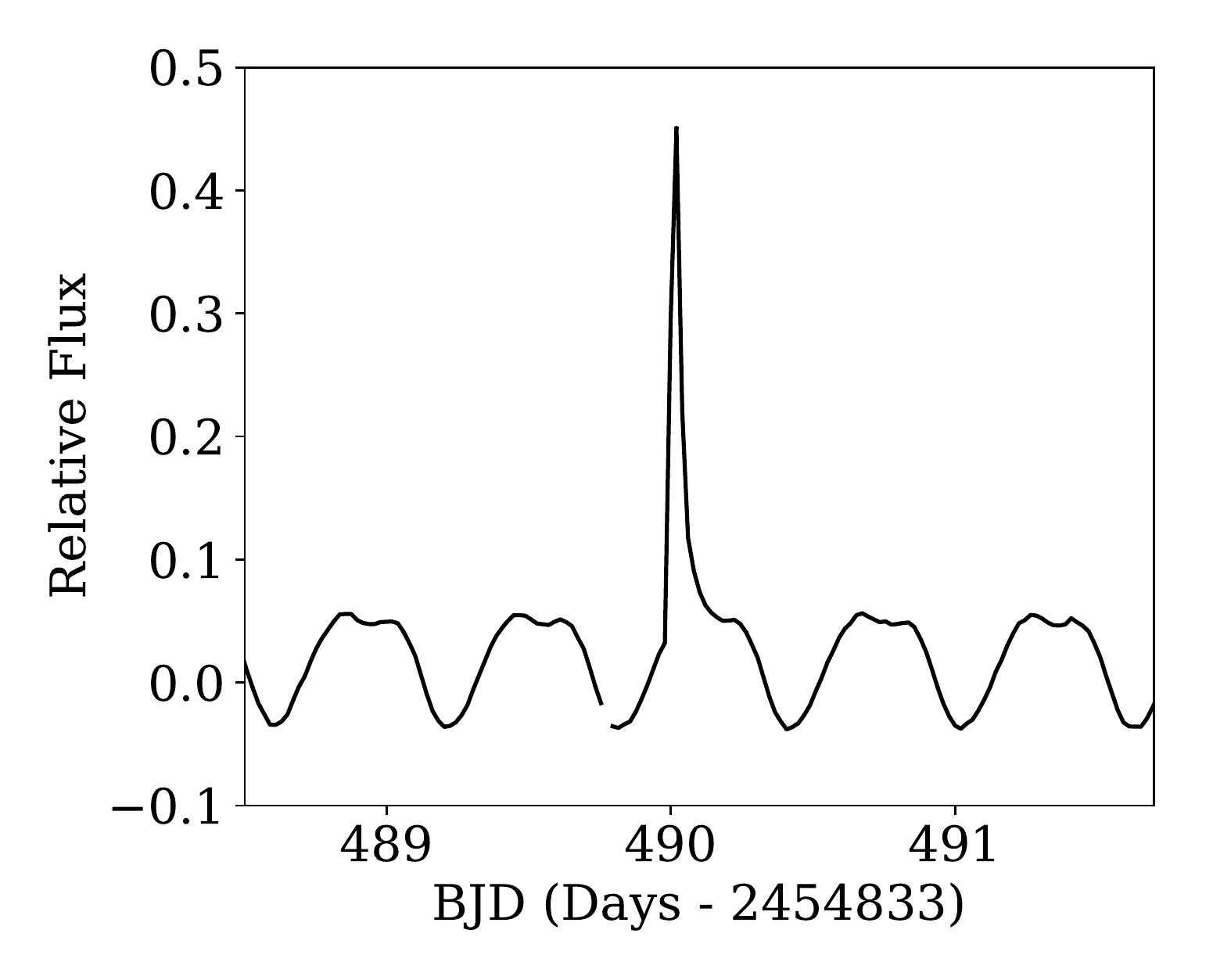}
\includegraphics[width=2.25in]{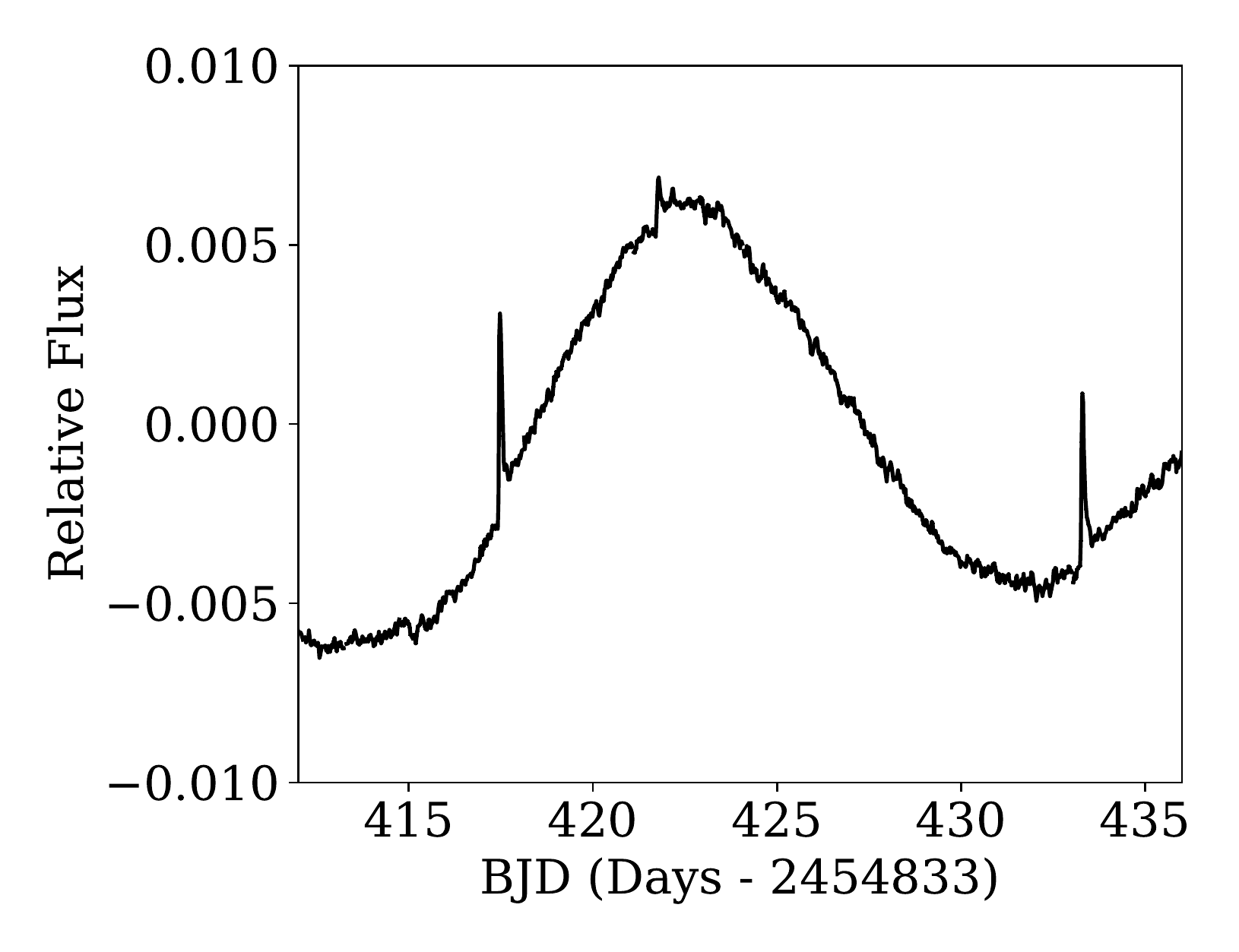}
\includegraphics[width=2.25in]{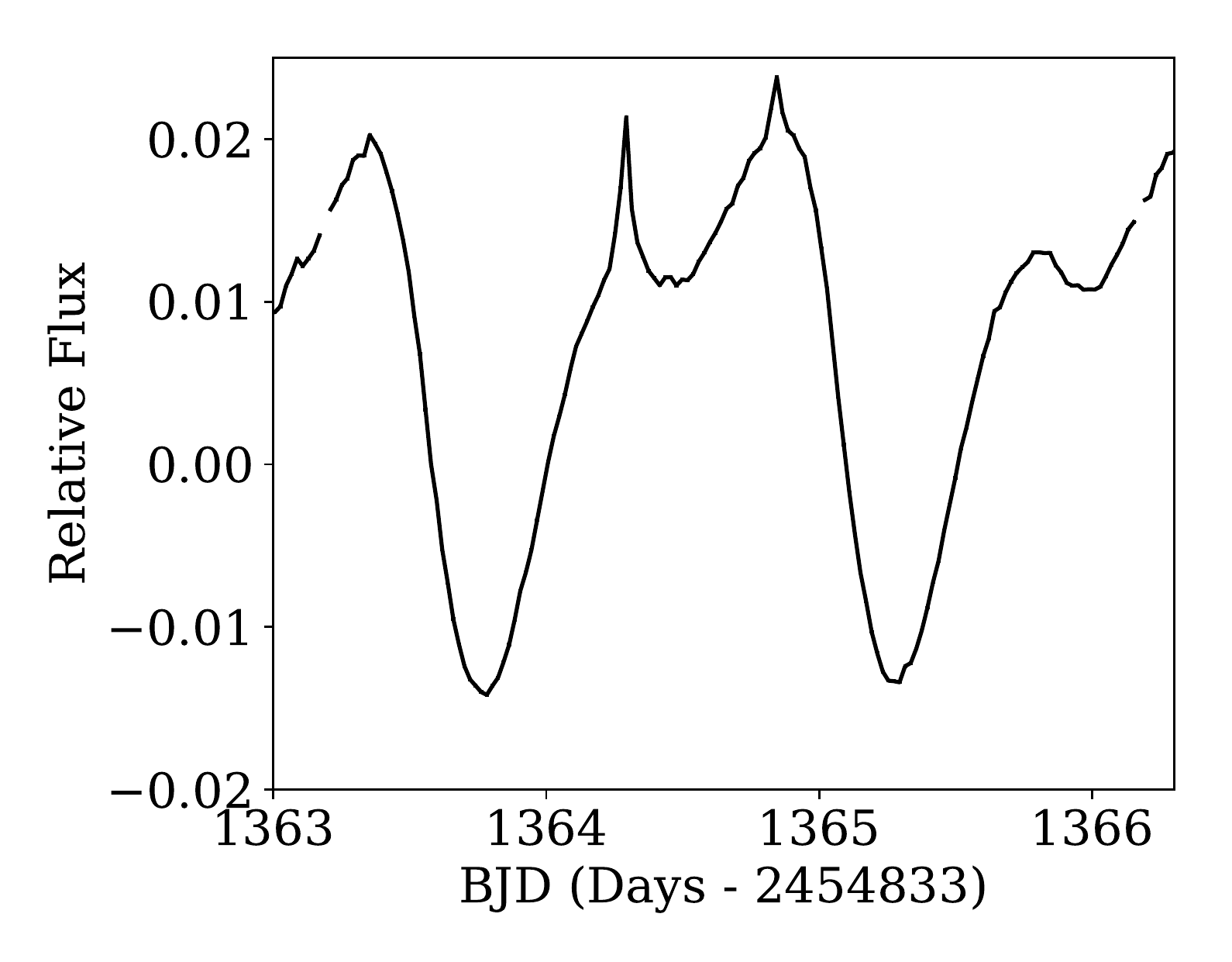}\\
\includegraphics[width=2.25in]{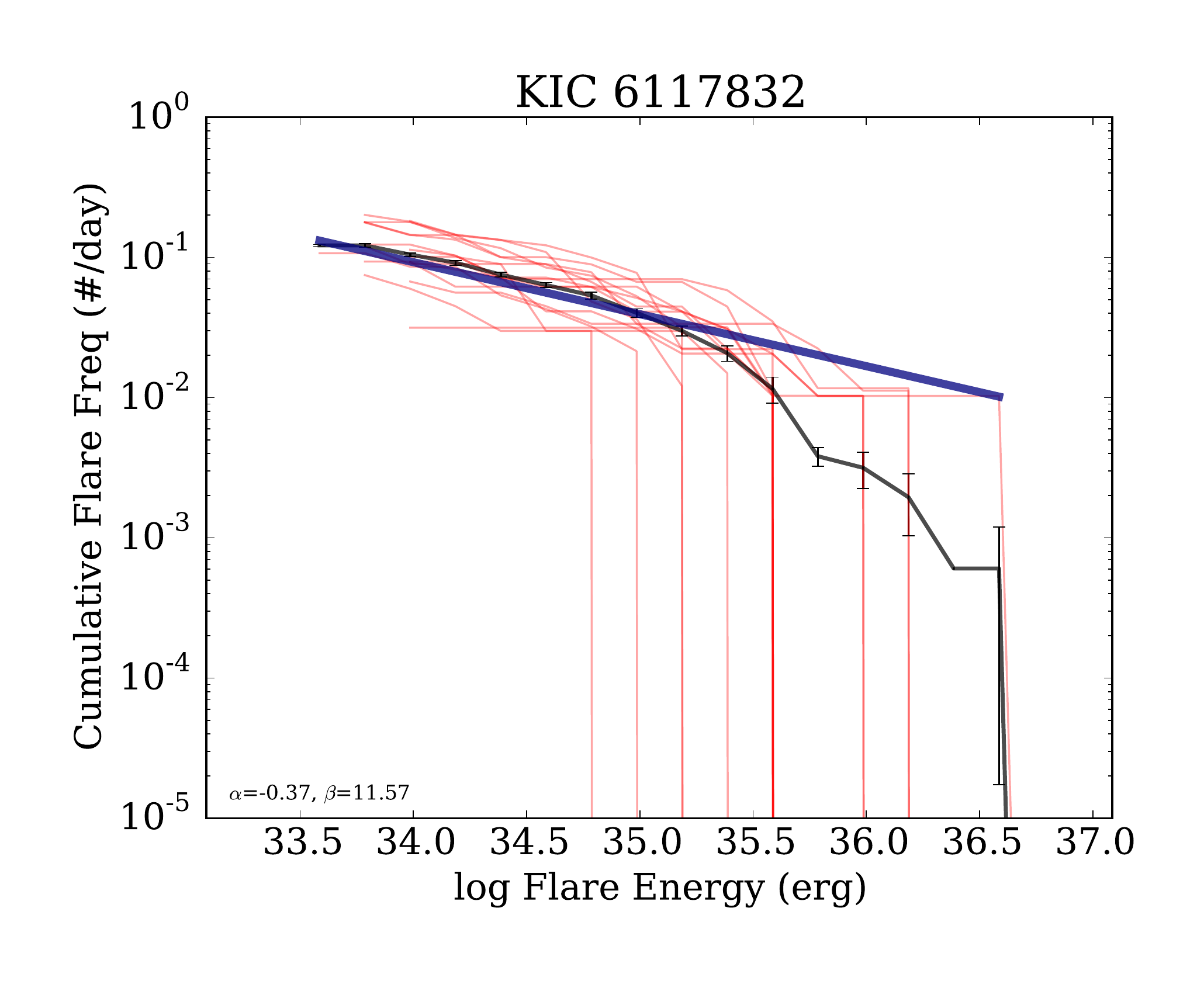}
\includegraphics[width=2.25in]{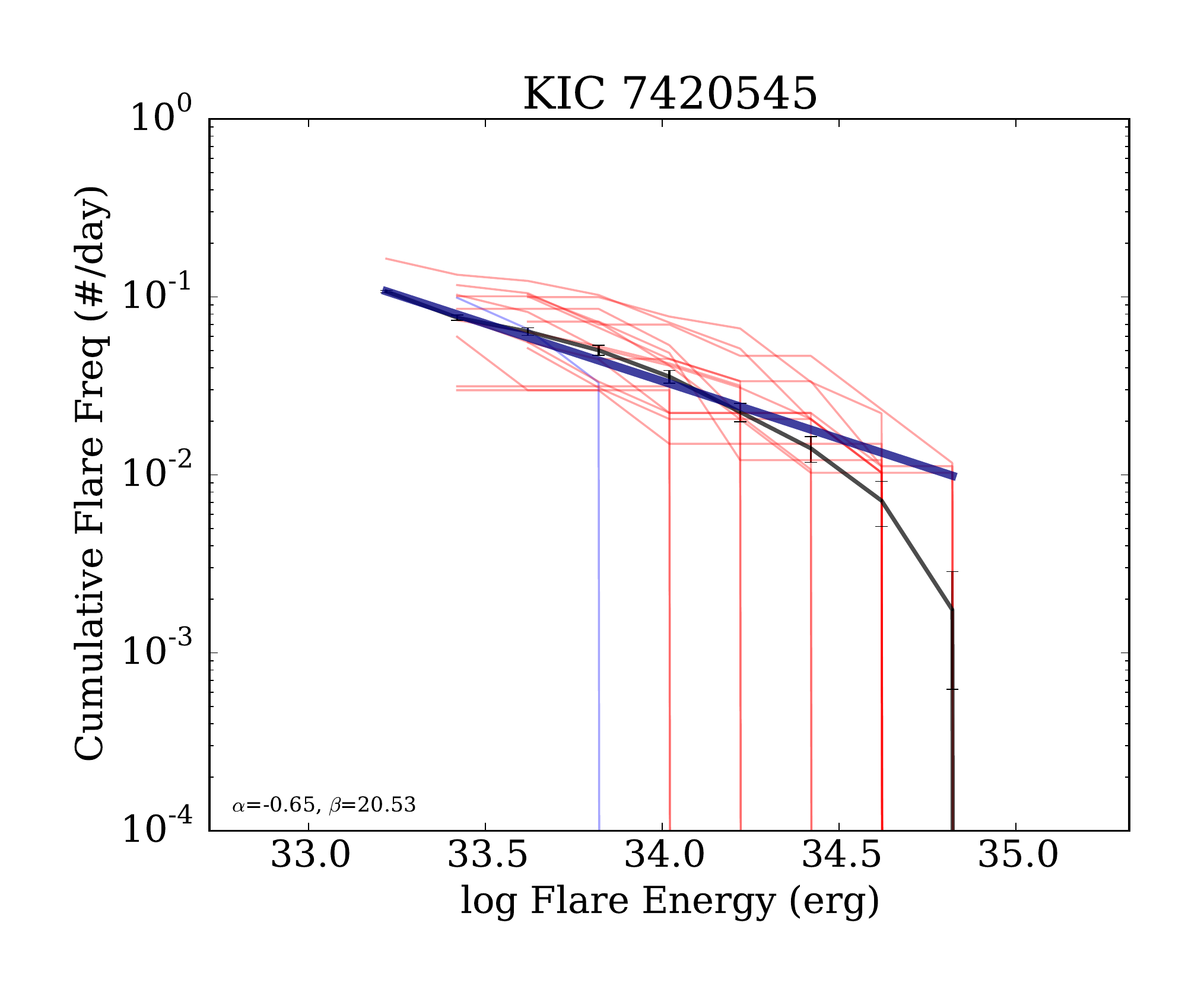}
\includegraphics[width=2.25in]{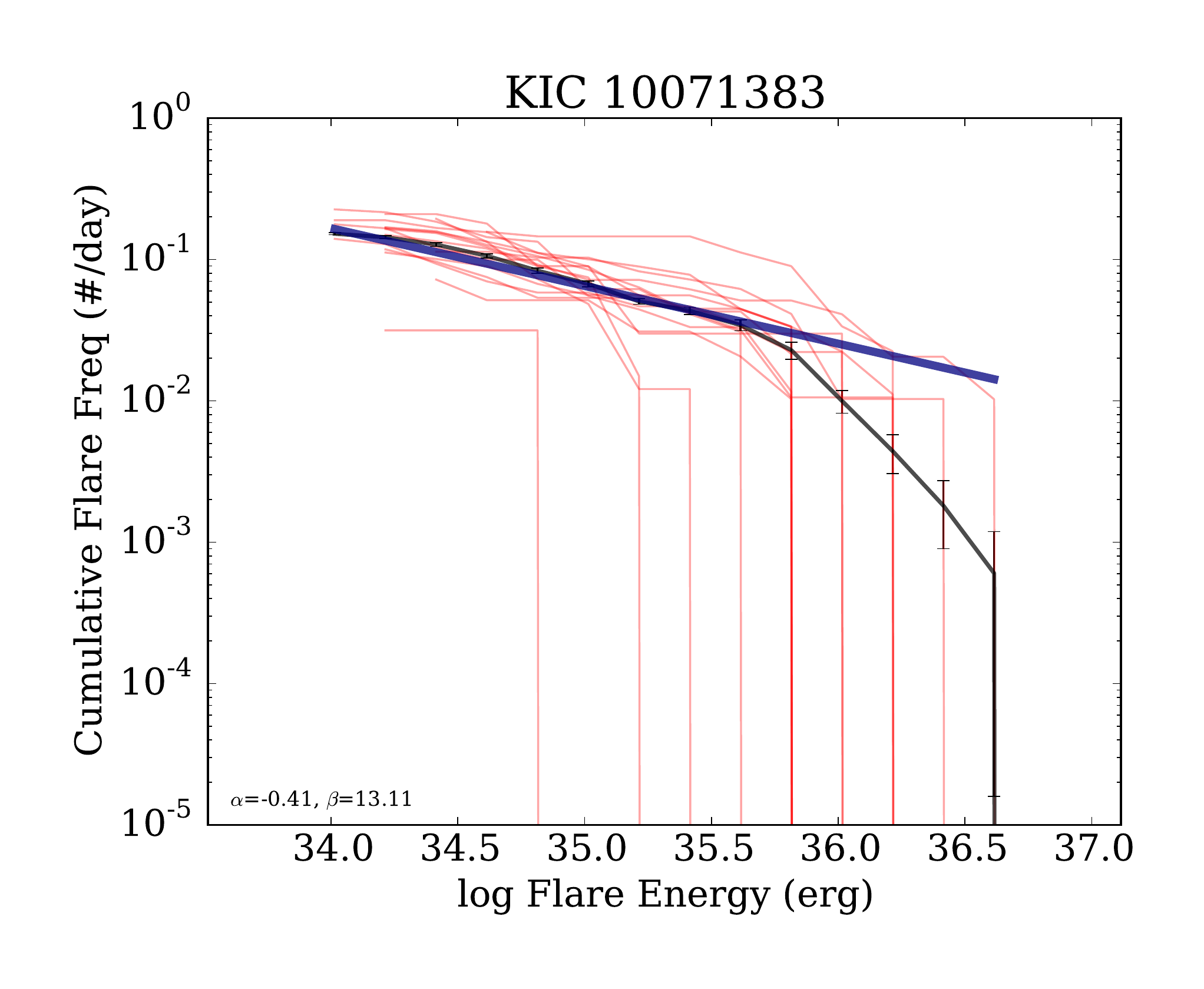}
\caption{
Three examples of flare stars from the \citet{davenport2016} sample. top row: sample light curves. Bottom row: cumulative flare frequency distributions (FFDs) from the{\tt appaloosa} flare finding analysis of \citet{davenport2016} for the same three stars. The FFD from each short cadence (blue lines) and long cadence (red lines) dataset is included, as well as the mean FFD (black) with Poisson uncertainties shown. Each star's mean FFD is fit with a power law (heavy navy line), whose slope and intercept in log--log space ($\alpha$ and $\beta$) are noted.
}
\label{fig:ffd1}
\end{figure*}

In Figure \ref{fig:ffd1} we show light curves and cumulative flare frequency distributions (FFDs) for three example flare stars from the \citet{davenport2016} sample that also have rotation periods measured in \citet{mcquillan2014}. These example stars will be followed through the analysis of this paper, and were selected to demonstrate a range of flare rates and rotation periods. 
Each star selected for inclusion in the \citet{davenport2016} sample had, within all available \Kepler data, at least 100 candidate flares of any energy, and 10 flares with energies above the 68\% completeness threshold determined by automated flare injection tests. These conservative thresholds potentially eliminate real flare stars with fewer significant flare events from our analysis, as noted by \citet{van-doorsselaere2017}, and may bias our sample towards active (younger) stars. 
In total we study the flare rate evolution from 347 stars with rotation periods from \citet{mcquillan2014} that pass the \citet{davenport2016} selection criteria for sufficient numbers of recovered flare event candidates. We use this conservatively selected, robust sub-sample of 347 stars from the \citet{davenport2016} catalog throughout this paper to explore in detail various approaches to quantifying and comparing flare activity as a function fo stellar age and mass.

\subsection{Flare Stars Among the Kepler Field}

\begin{figure}[!ht]
\centering
\includegraphics[width=3.5in]{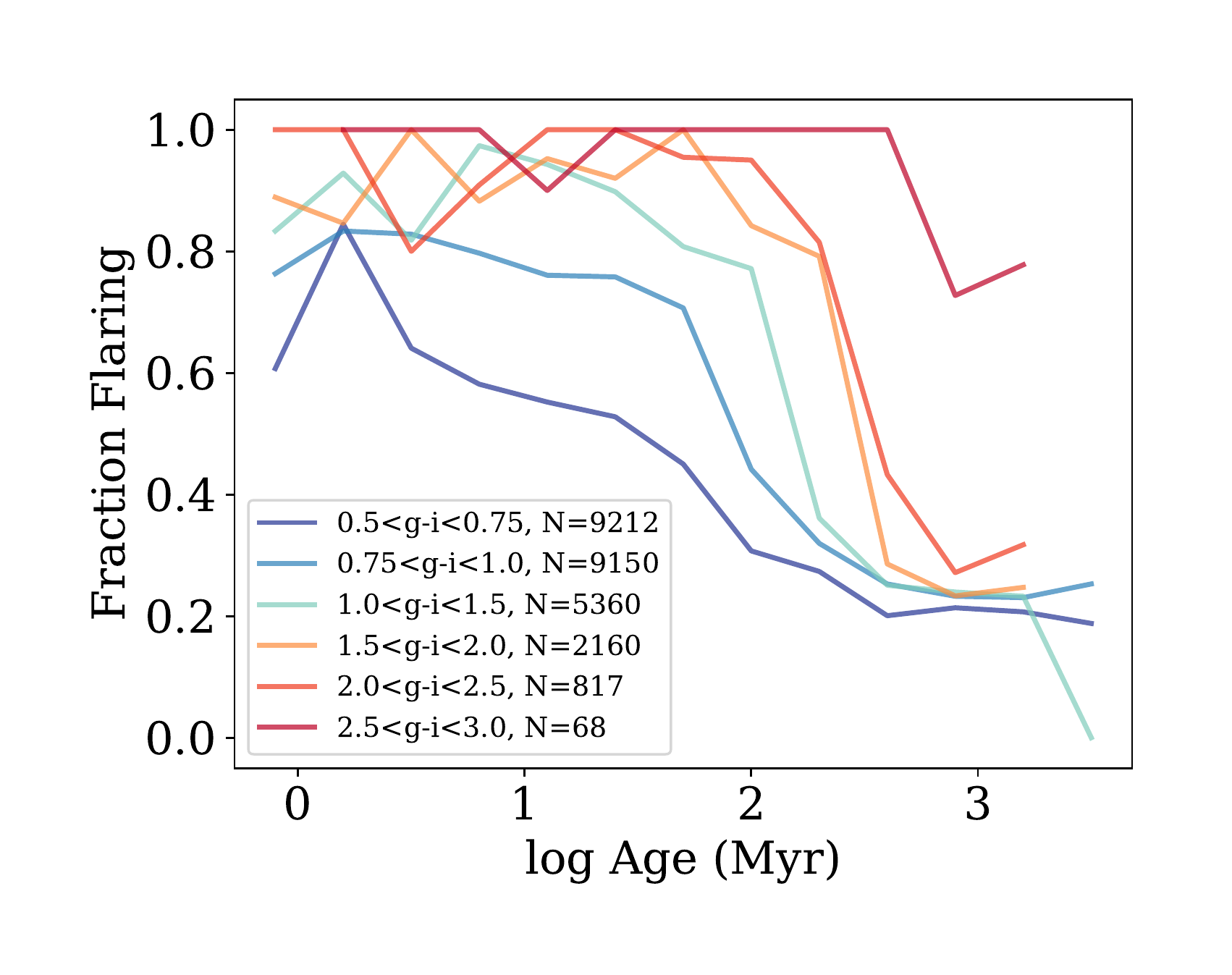}
\caption{
Fraction flaring stars as a function of their gyrochronology ages for 27,127 stars in six bins of $g-i$ color. The number of stars within each color bin is noted in the legend.
Each star has a rotation period determined by \citet{mcquillan2014}, and had artificial flare injection tests run by \citet{davenport2016}. Flaring stars were selected here as having at least 3 flare events above the 68\% recovery completeness threshold in \citet{davenport2016}. We find a trend of increasing total fraction of active flare stars, and an increasing apparent lifetime for flare activity, with decreasing mass (redder stars).
}
\label{fig:fracactive}
\end{figure}

As an aside, the thresholds for flare star selection from \citet{davenport2016} can instead be {\it relaxed} to consider the activity from a larger sample of stars, and to better explore the incidence of flare activity in the population of \Kepler field stars. As a demonstration of this, in Figure \ref{fig:fracactive} we show the fraction of flaring stars found in six bins of $g-i$ color as a function of their gyrochronology age. In this example we analyzed 27,127 stars from \citet{mcquillan2014} that had rotation periods between 0.1 and 30 days, and colors redder than $g-i \ge 0.5$ mag. Stars were considered ``active'' here if they had at least 3 flare events above the 68\% completeness threshold determined by the automated flare injection tests from \citet{davenport2016}, and had no requirement for the total number of flare event candidates. This figure emulates the H$\alpha$ activity fraction work typically done for M dwarfs, which shows an increasing lifetime of magnetic activity for stars with decreasing mass \citep[e.g.][]{west2008}. Previous studies such as \citet{kowalski2009} and \citet{hilton2010} have explored the fraction of flaring M dwarfs as a function of their height above the Galactic disk (a dynamical proxy for age), and have found flare activity decreases at earlier ages than H$\alpha$ emission activity. 

The lower mass (redder) stars in Figure \ref{fig:fracactive} exhibit both a higher overall fraction of flare activity at all ages, and a qualitatively longer lifetime of flare activity. Note this sample has not been cleared of contaminants from e.g. subgiants, which have been shown by \citet{davenport2017} to significantly contaminate the rotating G dwarf sample from \citet{mcquillan2014}, and which may lower the overall flaring fraction for bluer stars plotted here.  A more detailed census of stellar rotation periods and ages for all \Kepler and K2 stars \citep[e.g. see][]{van-saders2018}, as well as an exploration on the robustness of identifying flare stars from small numbers of events is needed to fully understand these flare activity lifetimes.
Still, we believe this is the first demonstration of flare activity fractions (directly related to activity lifetimes) for G, K, and M field dwarfs together, and warrants further study.

\section{Quantifying Flare Activity}
\label{sec:activity}

To accurately measure the evolution of flare rates, we must find a suitable metric to characterize flare activity for an ensemble of stars. Typical magnetic activity indicators, such as H$\alpha$ or X-ray flux, are used as disk-integrated measures of magnetically-driven emission from the chromosphere or corona. These quantities are often presented as relative luminosities, normalized either to a continuum flux, or to the stellar bolometric luminosity.  
While the overall rate and maximum intensity of a given star's flares appear related to the star's global magnetic field strength, the specific properties of individual flare events (e.g. duration, amplitude, morphology) are dependent on small-scale magnetic active regions. Since flare energy is inversely proportional to the event occurrence frequency, the flare properties measured for a given star will also be dependent on the photometric depth and temporal baseline of the observations. Our flare activity metric must therefore represent the integrated properties of many individual flare events to accurately model a star's magnetic activity state.

In this section we outline three methods for quantifying the photometric flare activity between stars, as well as the merits and challenges of each approach. These metrics include: 1) the fractional energy emitted in the \Kepler band by flares ($L_{fl}/L_{Kp}$), 2) the cumulative flare rate evaluated at a specific energy, and 3) an analytic model of the entire flare frequency distribution (FFD). While all three of these metrics have utility, we believe the latter will be of most value to future investigations, for example in studies of planetary habitability and atmosphere evolution. 

We note that the pedagogical discussion here of comparing flare metrics between \Kepler stars is similar in many ways to the review on flare activity by \citet{kunkel1975}. This excellent review explored two different methods for quantifying and comparing flare activity from heterogeneous photometric studies: the integral of the light curve, and the rate at a given specific energy level, which are directly analogous to the first two approaches advanced here. The third approach, comparing the entire FFD between stars, also has a long history \citep[e.g. see Fig. 17 of][]{lme1976}, and has even been used for comparing \Kepler flare stars \citep[e.g.][]{ramsay2013,hawley2014}.

\subsection{Fractional Flare Luminosity}
\label{sec:fracL}

An intuitive metric for quantifying magnetic activity strength via flares is the total luminosity emitted by flares relative to the nominal quiescent stellar luminosity, written originally as $L_{fl}/L_{Kp}$ by \citet{lurie2015}. This quantity is inspired by traditional magnetic activity measures for low-mass stars, such as $L_{H\alpha}/L_{bol}$  \citep{walkowicz2004} or $L_X/L_{bol}$ \citep{pallavicini1981}, which are normalized relative to the bolometric stellar luminosity. In this case, as presented in \citet{lurie2015}, the flare luminosity is normalized to the star's quiescent luminosity, but {\it only} in the \Kepler bandpass. This metric was used by \citet{davenport2016}, and also recently by \citet{yang2017}.

The relative flare luminosity has many advantages as a magnetic activity strength indicator. First, it is algorithmically simple to compute by taking the integral of the flaring portion of the light curve, as described by \citet{kunkel1975}. For \Kepler light curves, this is done by de-trending the non-flaring (quiescent) light curves, including starspot variations, normalizing them to their average relative flux, and then integrating all the identified flares. Integrating the relative flux of a single flare results in a quantity known as the ``equivalent duration'' \citep[e.g. see][]{huntwalker2012}, which has units of time (typically seconds). By integrating the relative flux from {\it all} flares in a \Kepler light curve, we would again have units of time, and so as defined by \citet{lurie2015}, $L_{fl}/L_{Kp}$ is simply computed as the integral of the relative flux of all flares, divided by the total observation duration of the light curve, resulting in a unit-less ratio.

A second appealing aspect of $L_{fl}/L_{Kp}$ as a flare activity indicator is that it compresses the entire observed flare activity of a star, regardless of the duration of the observation window, into a single number. This results in a quantity that has higher signal-to-noise than a specific flare rate, for example. This makes $L_{fl}/L_{Kp}$ ideal for comparing flare activity between stars, even with different observing baselines (e.g. comparing flare activity between active stars with differing numbers of quarters observed by \Kepler).

Thirdly, the light curve data does not need to be flux calibrated, or have accurate distances determined to measure $L_{fl}/L_{Kp}$. Instead the metric is defined totally by the relative flux increases of the flares. This is especially useful for datasets like \Kepler, where the light curves have incredible short-term precision designed to detect small amplitude exoplanet transits, but suffer from large-scale systematics that typically prevent flux calibration. As in the exoplanet transit application, the \% change of the light curve is the only quantity required. This also results in $L_{fl}/L_{Kp}$ being easily compared for many stars at once. \citet{lurie2015}, for example, used $L_{fl}/L_{Kp}$ to measure flare activity for the M5+M5 binary system GJ 1245 AB, and to relate this flare activity to other mid-to-late type M dwarfs such as GJ 1243 \citep{davenport2014b}.

\begin{figure}[!t]
\centering
\includegraphics[width=3.5in]{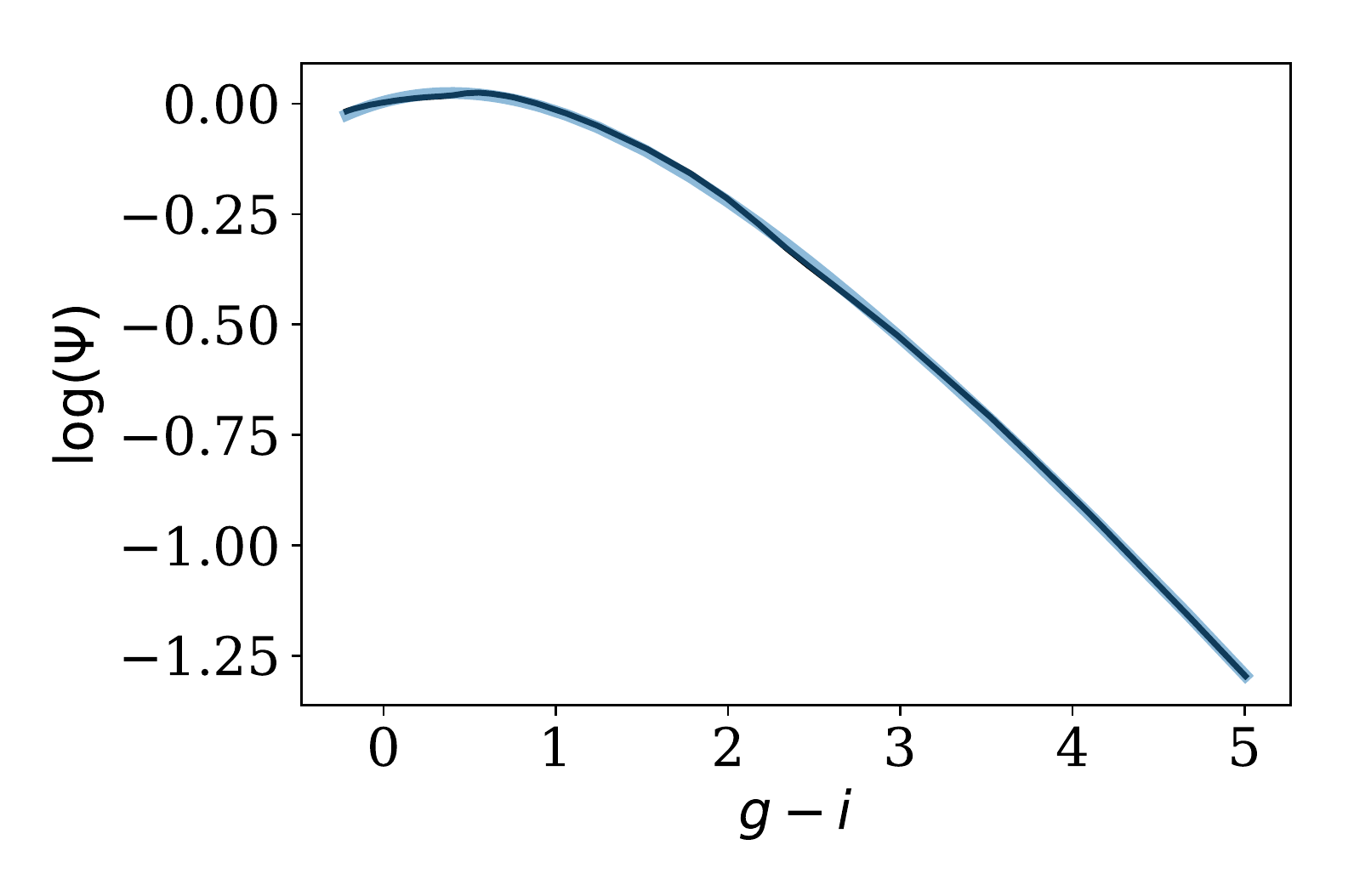}
\caption{
Flare $\Psi$, the correction factor (black line) that can be used to convert the observed fractional flare energy $L_{fl}/L_{Kp}$ in to the metric for comparing between stars of different masses, $L_{fl}/L_{bol}$. This $\Psi$ was computed using the bolometric and \Kepler absolute magnitudes from a 600Myr PARSEC isochrone \citep{bressan2012}. Here we show this correction factor versus $g-i$ color as a proxy for mass or spectral type \citep[e.g. see][]{covey2007,davenport2014}. A polynomial fit to the $\Psi$ factor is also shown (blue line), and is described in the text below.
}
\label{fig:chi}
\end{figure}

As \citet{lurie2015} note, to correctly compare flare activity between stars of varying spectral types (or effective temperatures), the measured quantity $L_{fl}/L_{Kp}$ requires a bolometric luminosity correction. Specifically, a correction must be made for the varying portion of the bolometric flux that is observed within the \Kepler bandpass. This is akin to the ``$\chi$'' parameter, first developed to convert H$\alpha$ equivalent width measurements into  $L_{H\alpha}/L_{bol}$, developed by \citet{walkowicz2004}. \citet{douglas2014} recently produced a thorough discussion on developing a $\chi$ factor using model spectra, and produced an updated table of $\chi$ values as a function of photometric colors in many bands. In extending this concept to the calculation of the $L_{fl}/L_{bol}$ ratio, we used the letter $\Psi$ as it follows $\chi$ in the Greek alphabet.

In Figure \ref{fig:chi} we demonstrate a similar parameter that can be used to convert measured $L_{fl}/L_{Kp}$ values into $L_{fl}/L_{bol}$, and thus more accurately compare the flare activity level between stars of different masses. The parameter $\Psi$ was determined using the \Kepler and bolometric luminosities computed for main sequence stars in a 600 Myr isochrone from the PARSEC model grid \citet{bressan2012}. The very wide bandpass of the \Kepler filter means the $\Psi$ factor is relatively close to 1 for most stars, and doesn't change much between spectral types. For ease of use, we also provide a simple polynomial fit to the curve shown in Figure \ref{fig:chi}:
\begin{eqnarray}
\log \Psi =& -0.0013 (g-i)^4 + 0.021 (g-i)^3 \notag \\
& -0.146 (g-i)^2 + 0.105 (g-i) \notag \\ 
& + 0.004
\end{eqnarray}

Note that the $\Psi$ parameter assumes a ``gray'' spectral response for the flare itself over the \Kepler bandpass. While H$\alpha$ is effectively emitted at a single wavelength, flares emit energy over all observed wavelengths. In optical wavelengths, the shape of this emission is typically characterized as a hot blackbody with $T_{eff}\approx10,000$ K, but has been observed to have significant excess emission in both the blue and red \citep[e.g. see][]{kowalski2013}. This effective temperature also changes throughout the flare event in a manner that is not well characterized, particularly for complex, multi-peaked flare events \citep[e.g.][]{slhadleo,kowalski2012}. \citet{davenport2016} discussed this in terms of estimating the energy of a single flare event, assuming a gray flare response across the \Kepler band. Some other studies assume a flare spectral model, which can in turn imply higher energies for specific events \citep[e.g.][]{gizis2013,maehara2015}. The $\Psi$ parameter shown in Figure \ref{fig:chi} will similarly underestimate the flare luminosity. However, as in \citet{davenport2016} we believe the best approach is to assume a gray response and not assume a single flare spectrum for all moments of every flare event.

\begin{figure}[!ht]
\centering
\includegraphics[width=3.25in]{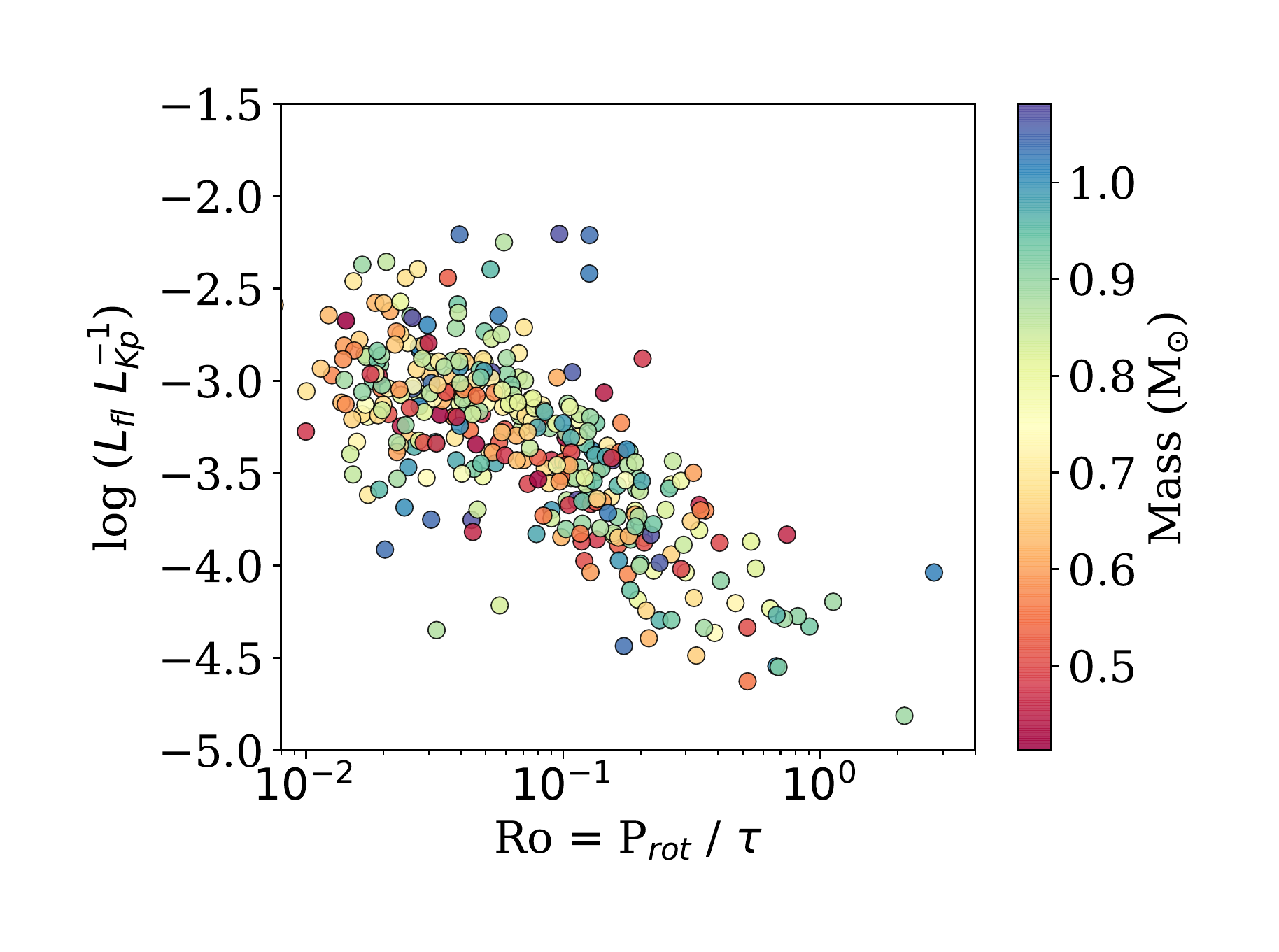}\vspace{-0.225in}
\includegraphics[width=3.25in]{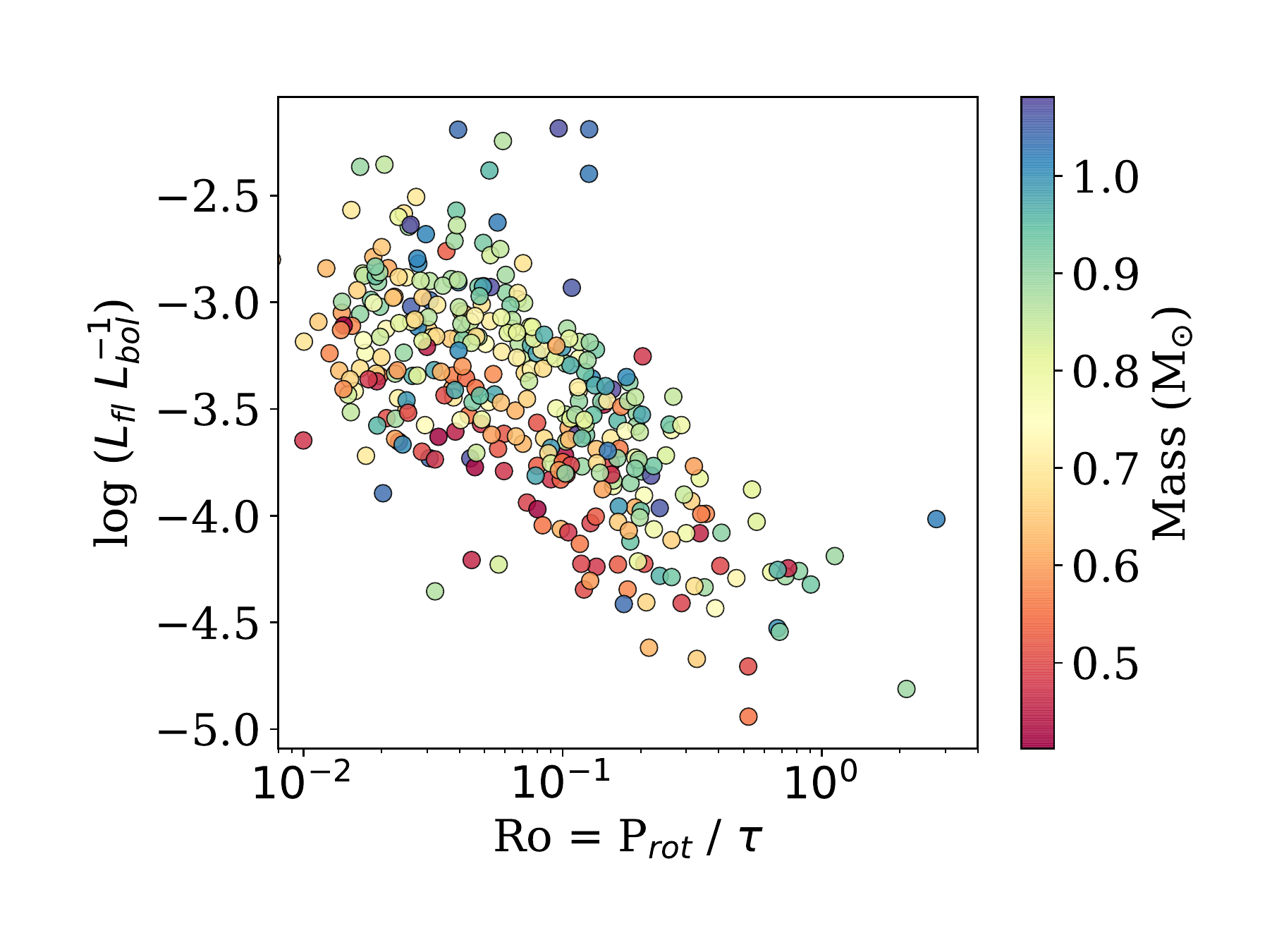}\vspace{-0.225in}
\includegraphics[width=3.25in]{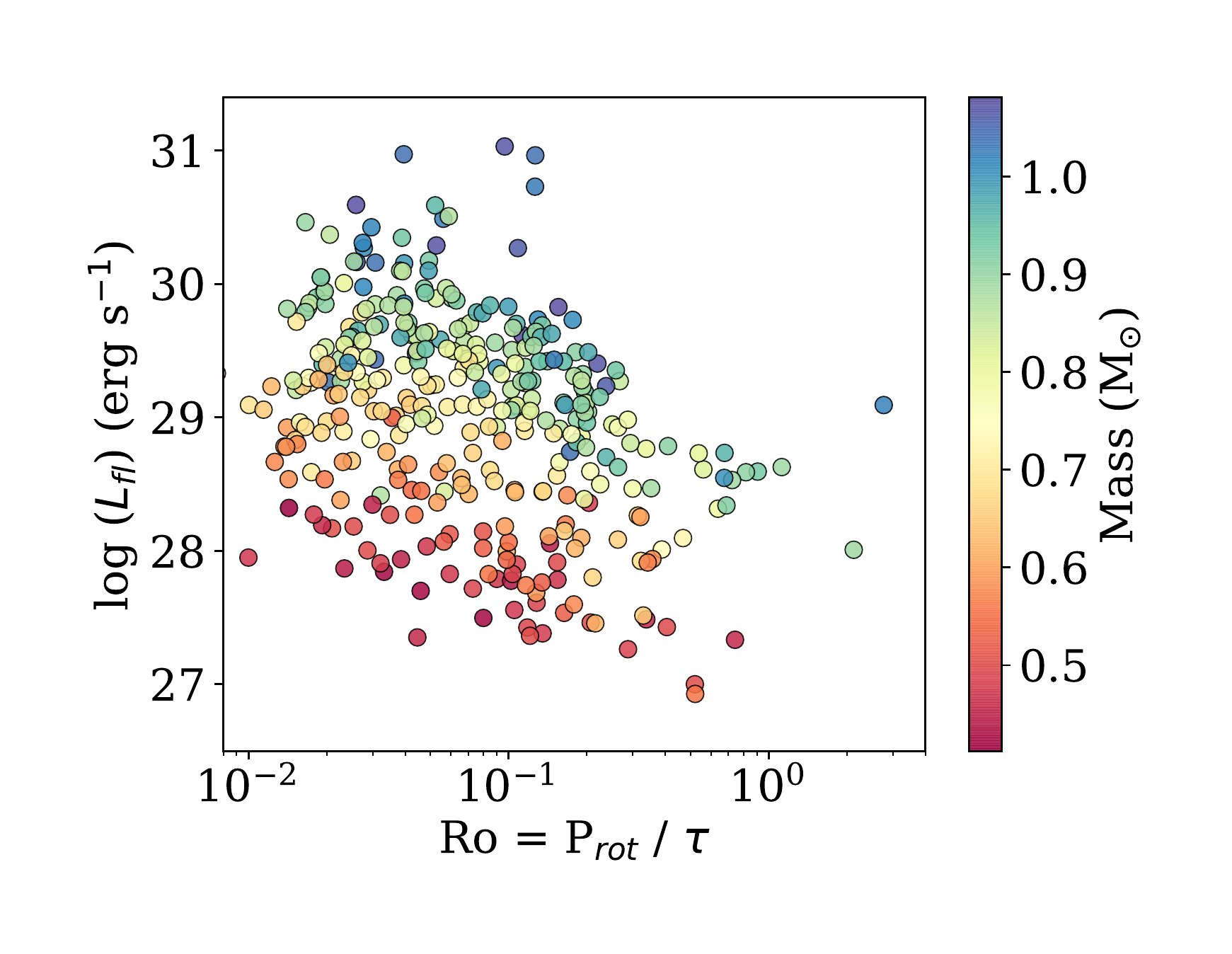}
\caption{
Rossby number vs flare energy for sample from Paper 1.
Top: The original $L_{fl}/L_{Kp}$ metric, as shown in \citet{davenport2016} versus Rossby number, with points colored by the stellar masses estimated by \citet{davenport2016} .
Center: The new $L_{fl}/L_{bol}$ metric, corrected using the $\Psi$ parameter, with point colors as above.
Bottom: Total estimated luminosity emitted in flares as a function of Rossby number, with point colors as above.
}
\label{fig:rossby1}
\end{figure}

\citet{davenport2016} found that the fractional flare luminosity, $L_{fl}/L_{Kp}$, decreased with increasing Rossby number, which is defined as the rotation period divided by the convective turnover timescale (Ro = P$_{rot} / \tau$). This result indicates that the flare activity is decreasing with stellar age, as expected from other tracers of magnetic activity. In Figure \ref{fig:rossby1} we reproduce the result of \citet{davenport2016}, coloring each point by the stellar mass. 
The center panel of Figure \ref{fig:rossby1} shows the $\Psi$ corrected relative flare luminosity, $L_{fl}/L_{bol}$. Since the $\Psi$ correction factor is near 1 for most stars, the change between these panels is modest. Interestingly, a slight gradient in the $L_{fl}/L_{bol}$ as a function of mass appears for very rapidly rotating stars, particularly those within the ``saturated dynamo'' regime (Ro $< 0.1$). The higher mass (near Solar mass) stars show a larger fraction of their luminosity emitted through flares in this regime. Though our sample of slower rotators is small, this gradient in $L_{fl}/L_{bol}$ seems to disappear for larger Rossby numbers. 

An additional complication worth mentioning in the use of the relative flare luminosity metric is due to the varying distances and luminosities of stars in a given magnitude-limited sample. Since detection of flare events (particularly the small-amplitude, lower energy events) depends on the signal-to-noise ratio of a light curve, the distances to stars can impact the resulting $L_{fl}/L_{Kp}$ measurement. 
For example, given two identical mass stars with the same underlying flare activity level placed at different distances, the more distant star's lower amplitude flares will be obscured by photometric noise, and thus returning a lower $L_{fl}/L_{Kp}$ measurement. Similarly, for two stars at the same distance, the lower-mass (fainter) one will have a lower signal-to-noise, and again the flare activity will be under-estimated. As a result, when comparing stars using $L_{fl}/L_{Kp}$ (or $L_{fl}/L_{bol}$), a uniform range of flare event energies must be considered. This presents a major limitation in comparing the flare activity between stars of different masses. We therefore generally recommend $L_{fl}/L_{Kp}$ only be used when comparing stars within a narrow mass range.

In Figure \ref{fig:rossby1} we also show the total luminosity emitted in flares as a function of the Rossby number, again with points colored by the estimated stellar mass. $L_{fl}$ here is computed as $L_{fl}/L_{Kp}$ times the quiescent luminosity in the \Kepler band ($L_{Kp}$), which is used in \citet{davenport2016} in converting individual flare events from the observed ``effective duration'' into energy. A strong gradient in total flare luminosity is seen as a function of mass (i.e. as a vertical color gradient in the lower panel of Figure \ref{fig:rossby1}).
The total flare luminosity ($L_{fl}$) is an estimator of the total magnetic activity strength in flares (over the \Kepler band). Like the fractional flare luminosities described above, this is useful in comparing flare activity between stars, particularly for considering e.g. the total incident flare flux an exoplanet might experience.

\subsection{Specific Flare Rate}
\label{sec:rate}

Rather than integrating the relative flux from all detected flares within a light curve, as outlined in \S\ref{sec:fracL}, flare activity can be expressed as a specific occurrence frequency. Since observable flares occur with a wide range of energies for a given star, such a frequency should be reported for flares of a given energy {\it or larger} (e.g. 10 flares per year with energies of $10^{32}$ erg or larger). This is equivalent to evaluating the flare frequency distribution (FFD; see Figure \ref{fig:ffd1}) at a given energy. Here we present this metric as $R_{32}$, with the subscript denoting the $\log$ energy that rate is evaluated at, and with units of cumulative number per day.

The specific flare rate has been used in some capacity for many years for comparing flare activity levels between stars \citep[e.g.][]{lme1976}, and was noted by \citet{davenport2016} as an effective metric for comparing flare activity levels between stars at different distances. Since the specific flare rate is generated from the FFD, it can in principle be evaluated at energies {\it not} observed for a given star by fitting and extrapolating a powerlaw function to the FFD. Note this implicitly assumes that the flare rate for a star is governed by a single power law at all energies of interest, which is not always supported by observations. For example, as \citet{davenport2016} highlight for KIC 11551430, a ``break'' in the FFD power law is observed for superflare stars at high event energies. 

Projecting the specific flare rate using the single FFD power law is very useful for comparing stars with different observing conditions, such as 1) stars with very different observing durations where one star may not have produced many large flares to compare to, or 2) comparing stars at significantly different distances where small amplitude flares from the fainter object are not detectable. What is required for projecting a flare rate to a new energy range, however, is a sufficient number of flares be observed to adequately measure the power law distribution in the FFD.

This specific flare rate metric is also appealing due to its easily understood units, i.e. number of flares per day, and can be useful when considering the impact of flares on other observable properties. For example, \citet{davenport2016b} report for Proxima Cen a specific rate for superflares of $R_{33}=8$ per year that may impact exoplanet habitability, and a rate of $R_{28}=63$ per day for events having amplitudes comparable to an exoplanet transit signal.

\begin{figure}[!t]
\centering
\includegraphics[width=3.5in]{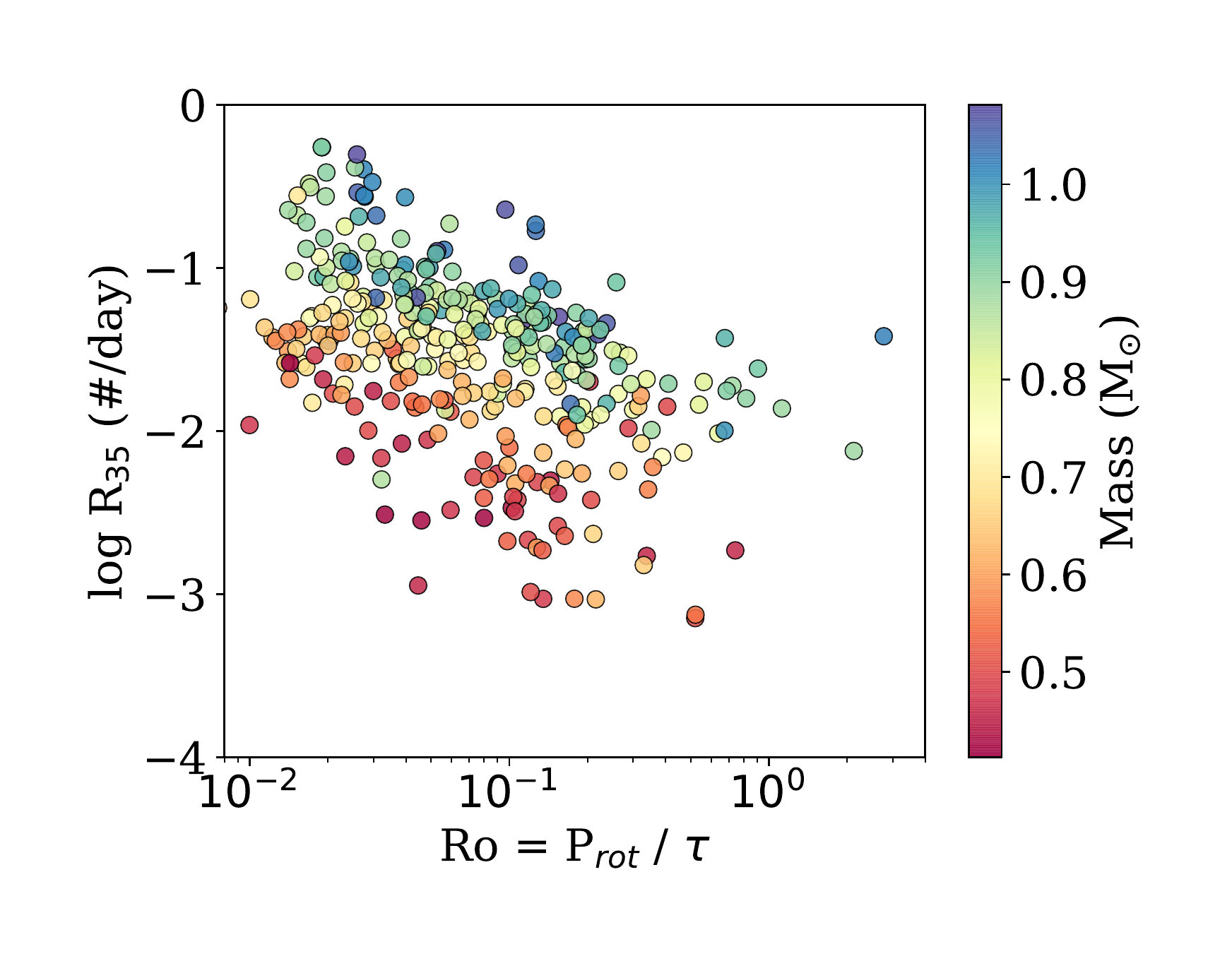}
\includegraphics[width=3.5in]{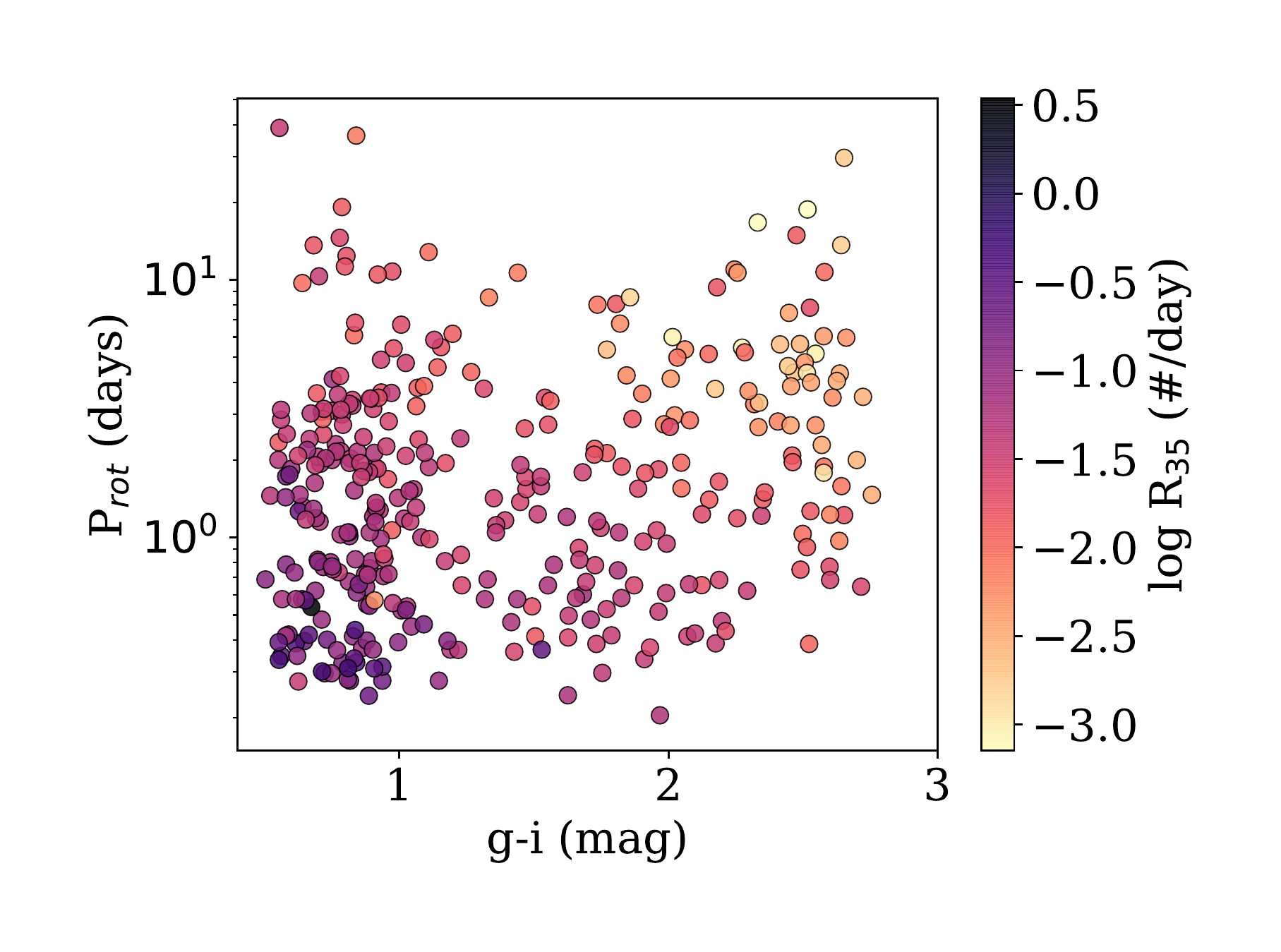}
\caption{
Specific flare rate ($R_{35}$), evaluated at an energy of log E = 35 erg as a function of the stellar mass and Rossby number (top), and the observed stellar rotation and color (bottom). The sample is the same as in Figure \ref{fig:rossby1}.
}
\label{fig:rossby2}
\end{figure}

In Figure \ref{fig:rossby2} we demonstrate the specific flare rate $R_{35}$ for the same sample of stars shown in Figure \ref{fig:rossby1}. Here we explore both the dependence on mass and Rossby number (top) and the observed color and rotation period (bottom). The energy of $10^{35}$ erg was selected as the average event energy observed in the \citet{davenport2016} flare census. For stars at a given mass or color the specific flare rate declines with increasing rotation period (or Rossby number), consistent with the results from the previous section. The specific flare rate is therefore a simple and effective way to compare flare rates between stars at different distances and under vastly different observing conditions, provided the stars are comparable in color (mass).

However, Figure \ref{fig:rossby2} reveals a gradient in $R_{35}$ as a function of either stellar mass or color, where lower mass stars have smaller specific flare rates at a given energy and rotation period (or Rossby number).
This manifests as an unintuitive trend where lower-mass stars appear {\it less} active as measured by their specific flare rate when compared to higher mass stars at a given rotation period.
This reproduces the previously known effect that while low-mass stars produce a high rate of observable flares and can produce a large fraction of their luminosity in flares, the actual rate of events with solar-type flare energies is low.

\begin{figure}[!ht]
\centering
\includegraphics[width=3.5in]{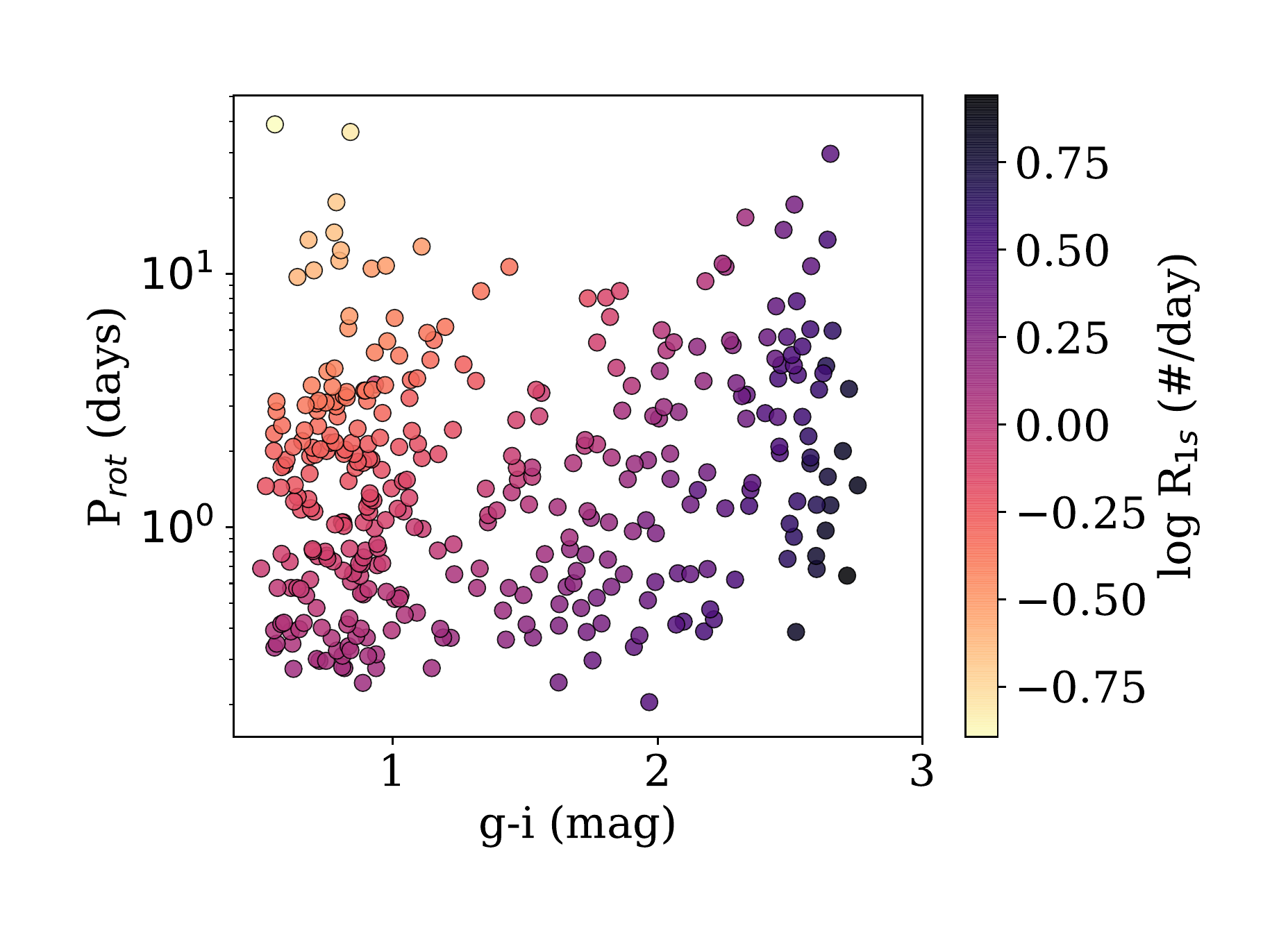}
\caption{
The Specific flare rate ($R_{1s}$), evaluated at an energy that is equal to an equivalent duration of 1 second
}
\label{fig:R1s}
\end{figure}

To overcome this, we can correct the specific flare rate for the difference in quiescent stellar luminosity between stars. Rather than pick a single event energy to evaluate all FFD's at, we can instead pick a flare energy that scales with the star's luminosity. A flare event with an equivalent duration of $P=1$ second has, by definition, an energy equal to the star's quiescent luminosity integrated for 1 second. A sensible energy to choose for evaluating the FFD would therefore be the star's quiescent luminosity in the \Kepler band\footnote{The bolometric luminosity would also be a good choice for defining the comparison energy, but as we explored in the development of the $\Psi$ correction factor, the ratio of $L_{Kp}/L_{bol}$ is typically near 1 for solar-type stars.}. We will denote this adjusted specific flare rate for events with an equivalent duration of 1 second as $R_{1s}$. In Figure \ref{fig:R1s} we show the rotation--color space for the same sample of stars as a function of their $R_{1s}$ specific flare rate. This adjusted specific flare rate exhibits the same decrease in flare activity with age (rotation), but the low-mass stars now display a higher flare activity level relative to their quiescent luminosity.

\begin{figure*}[!ht]
\centering
\includegraphics[width=2.25in]{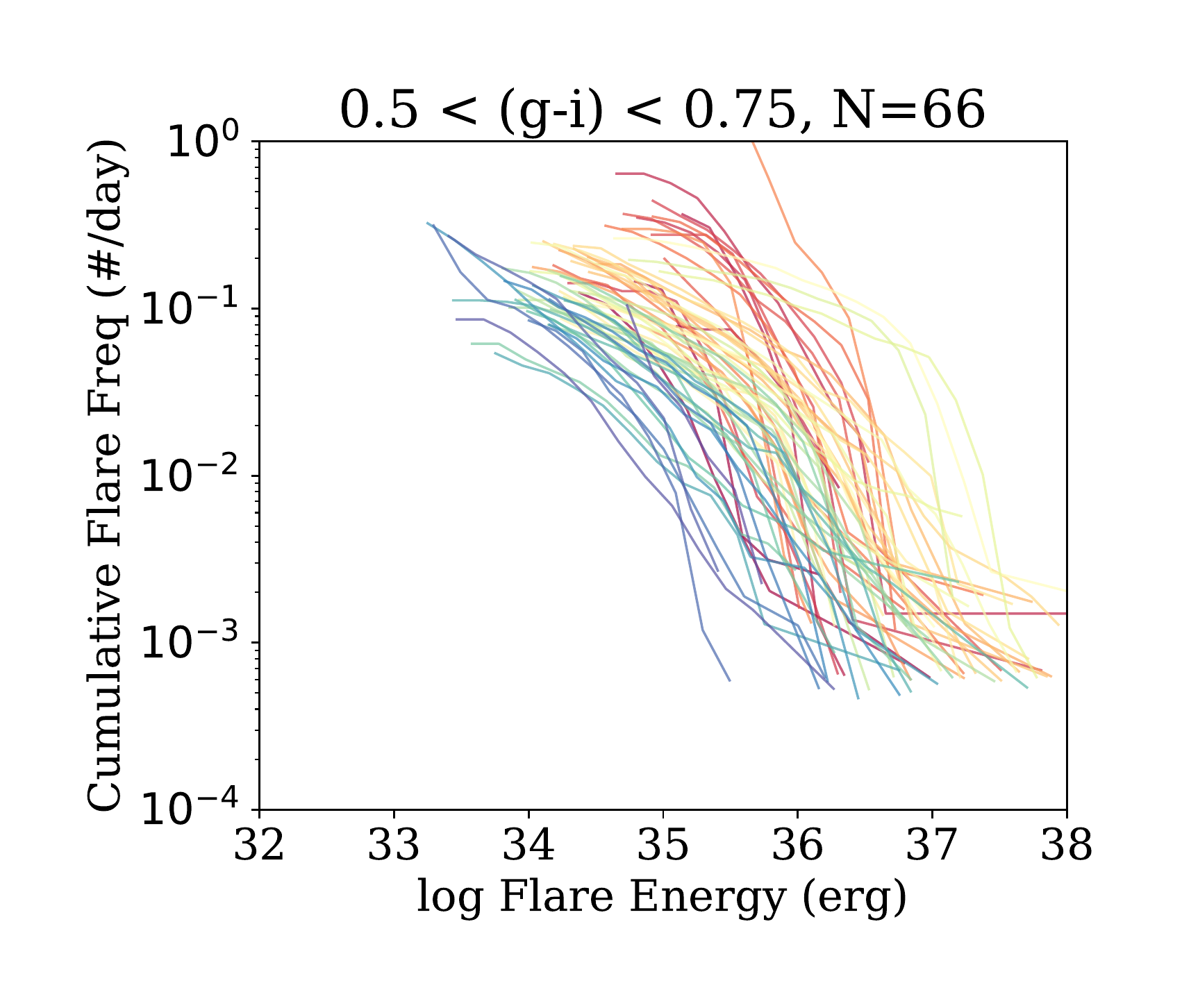}
\includegraphics[width=2.25in]{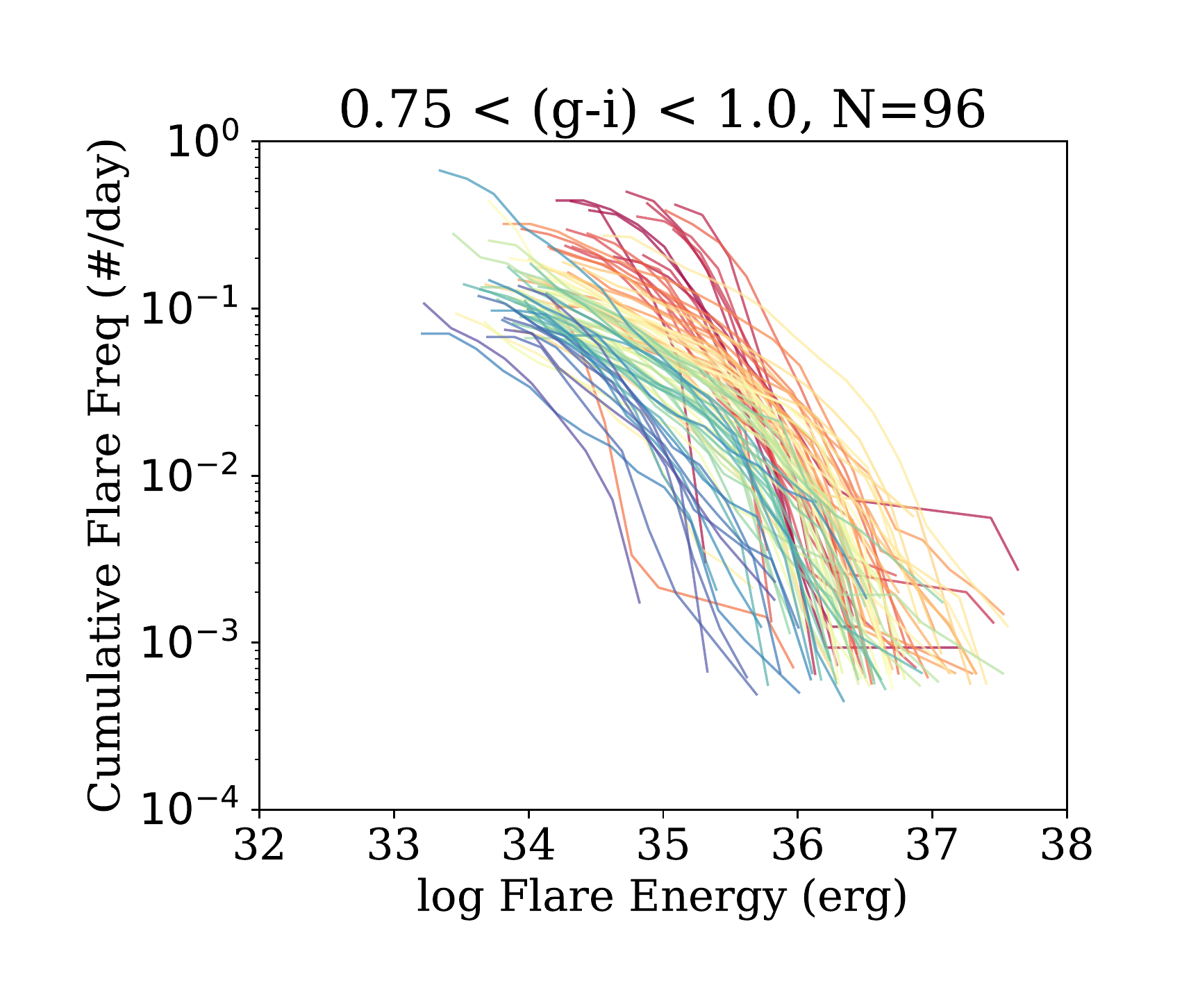}
\includegraphics[width=2.25in]{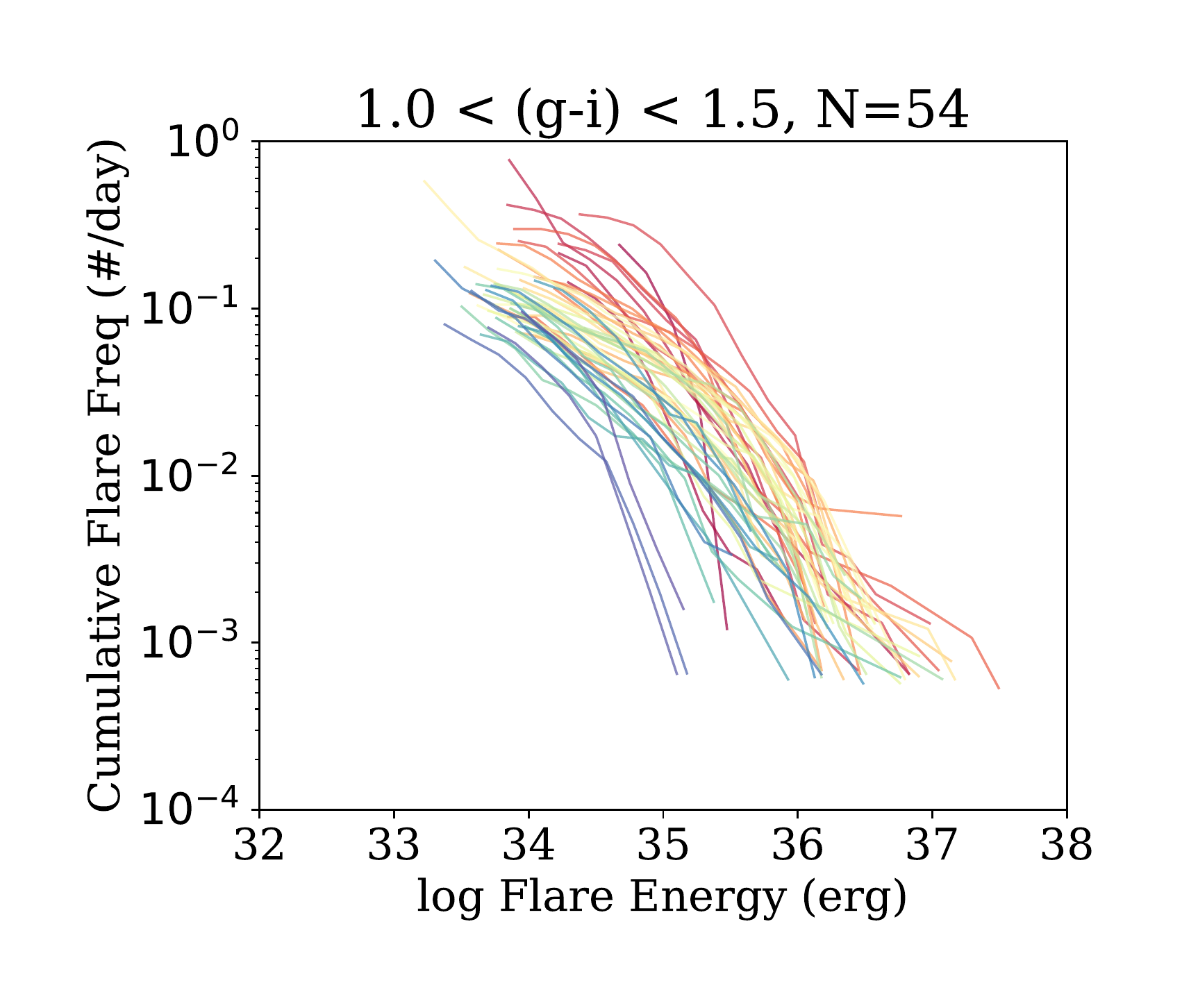}\\
\includegraphics[width=2.25in]{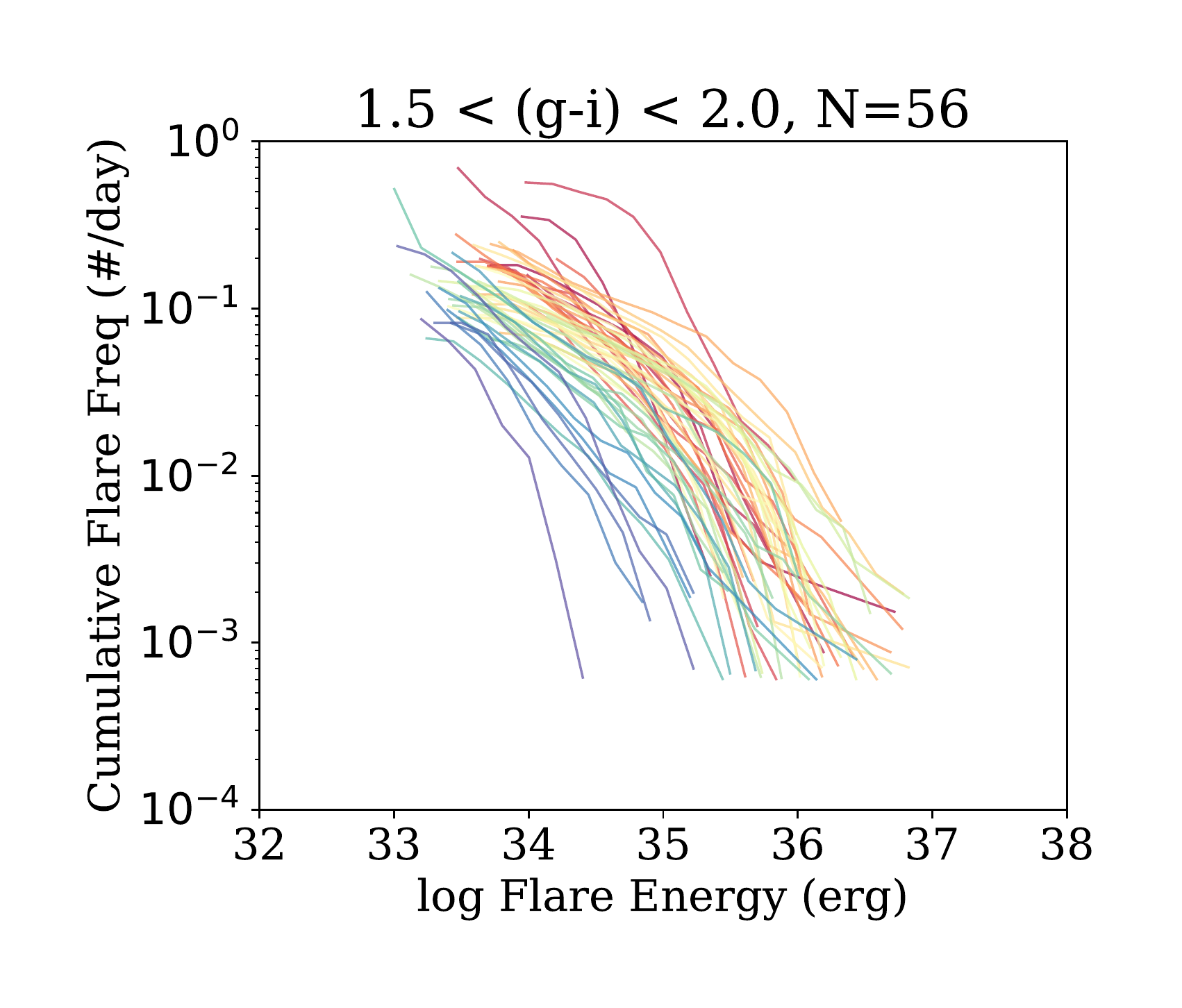}
\includegraphics[width=2.25in]{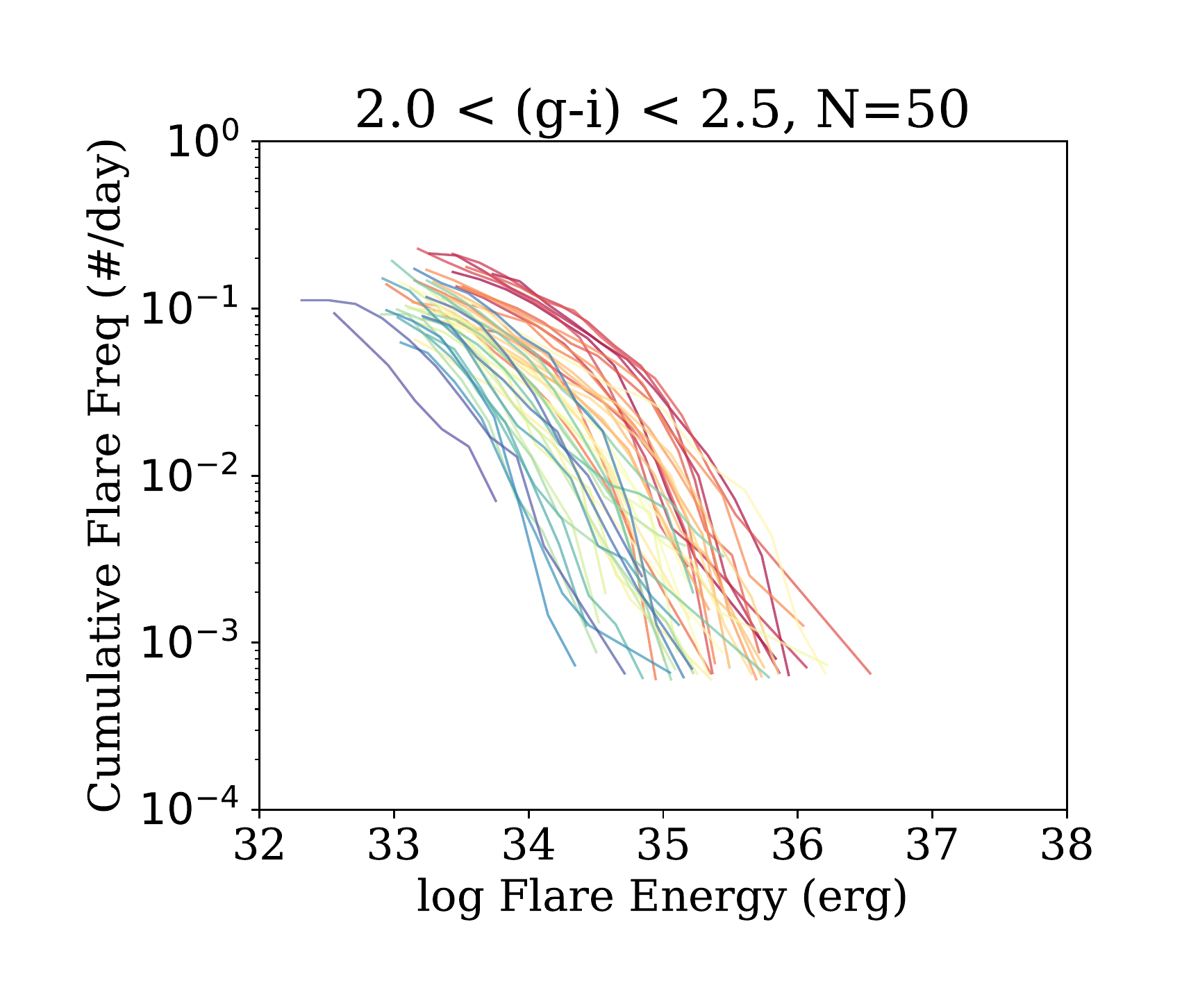}
\includegraphics[width=2.25in]{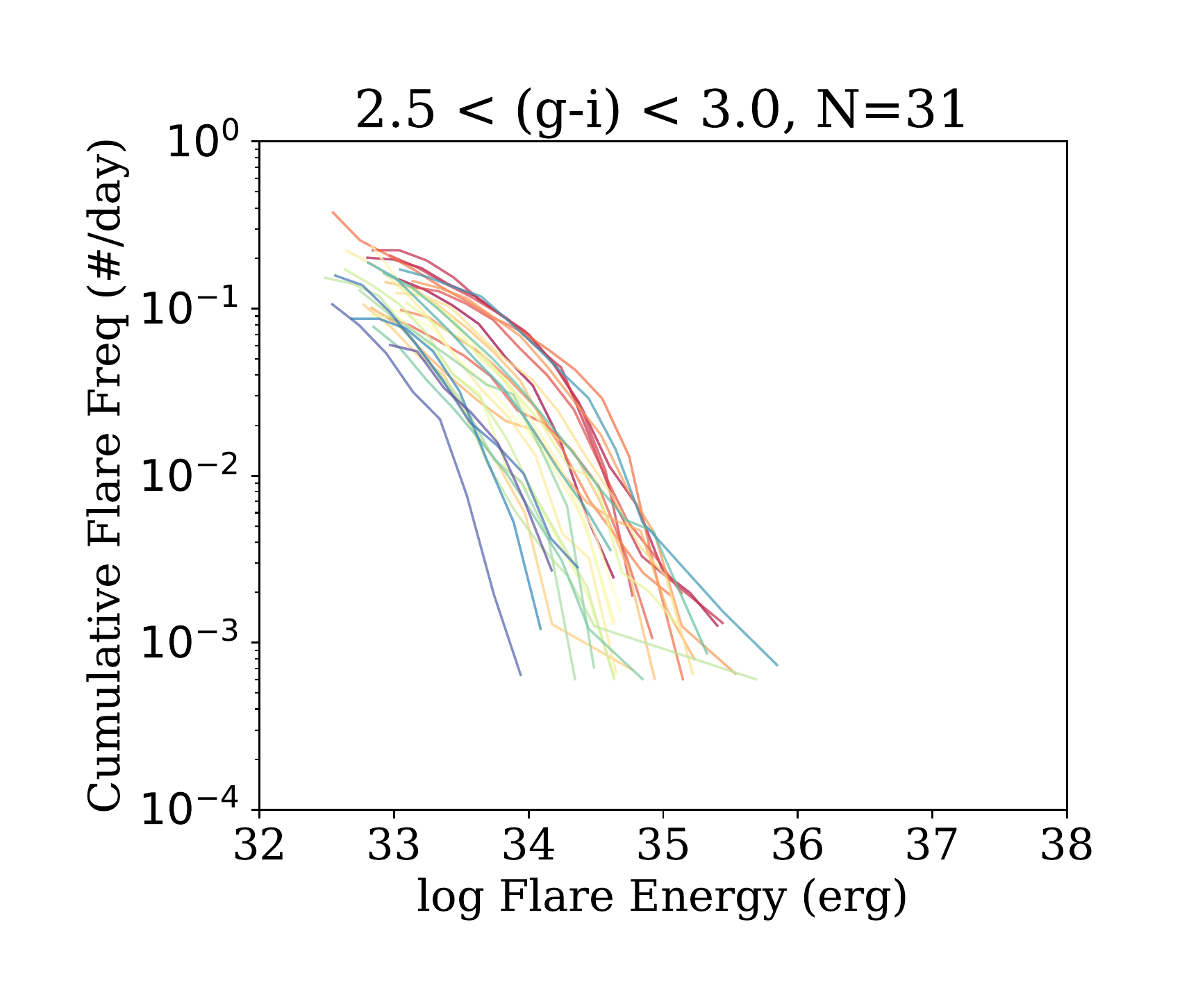}
\caption{
Average FFD, combining all available quarters of \Kepler data, for stars in the six $g-i$ color bins defined in \citet{davenport2016}. Each track is colored as a function of the measured rotation period from \citet{mcquillan2014}. A clear and coherent decrease in the total flare rates is observed as stars slow down (age) within each panel.
}
\label{fig:meanffd}
\end{figure*}

\subsection{Modeling the Flare Frequency Distribution}
\label{sec:ffd}

The previous two metrics presented ($L_{fl}/L_{bol}$ and $R_{1s}$) have the useful property of reducing the complexity of a star's observed flaring behavior into a single quantity. While these metrics are useful for easily comparing flare activity between stars (given the caveats discussed above), such simple metrics neglect the important structure that is present in the full FFD (flare frequency distribution). Rather than try to fully reduce the complexity of the FFD down to a single number, here we demonstrate how to quantitatively use the full FFD to describe the evolution of stellar flare activity.

Comparing evolution of the entire FFD between stars requires considering a four dimensional space: cumulative flare frequency (the ordinate variable) as a function of flare event energy, stellar mass (or color), and age (or rotation period). 
Figure \ref{fig:meanffd} shows the FFDs for our sample of 347 stars with measured rotation periods, separated into six bins based on their observed $g-i$ colors (as a proxy for mass).
Each line shows the mean FFD for a single star using all available quarters of \Kepler data, corresponding to the average FFDs shown as heavy black lines in the bottom panels of Figure \ref{fig:ffd1}.
Each FFD line is colored by the observed rotation period reported by \citet{mcquillan2014}, with  rotation periods increasing from red to blue. Thus the panels of Figure \ref{fig:meanffd} represent slices in mass of the four dimensional data space we are interested in (stellar mass, age, flare event energy, flare frequency). 

As presented in Section \S\ref{sec:rate}, the range of flare energies we can detect in each panel of Figure \ref{fig:meanffd} is affected by observational and astrophysical biases and limits.
The low-energy event cutoff for each star is determined by the automated flare injection and recovery tests described by \citet{davenport2016}. A bias is also seen where rapid rotators have slightly higher energy cutoffs due to increased starspot amplitudes that were not perfectly modeled out by {\tt appaloosa}, which made flare recovery more difficult for the iterative algorithm of \citet{davenport2016}.

Figure \ref{fig:meanffd} shows, in every color bin, a decrease in the observed flare frequency as a function of rotation period. This monotonic decrease in flare rate as stars lose angular momentum is the evolution we wish to model. Qualitatively, the flare activity decreases more dramatically for bluer stars, consistent with results from e.g. Figure \ref{fig:R1s}, and our expectations from other measures of stellar magnetic activity over time \citep[e.g.][]{shkolnik2014,nunez2017}. As \citet{davenport2016} highlighted for one flaring G dwarf, a ``break'' in the FFD power law is apparent at the highest flare event energies. This break energy also seems to scale with the energy range of events observed within each $g-i$ color bin (i.e. lower-mass stars have a lower break energy). Though we do not fully explore this feature here, we emphasize that it may represent an important limit on the maximum energy budget of stellar active regions, or their formation timescales as a function of the total stellar magnetic field strength.

\section{Analytic Flare Evolution Model}
\label{sec:ffdmodel}

To produce a general model for the FFD in physical units, we convert the observed colors and rotation periods for each star in our sample into masses and ages, respectively. The $g-i$ color for each star was converted to stellar mass using a 600 Myr isochrone, as described in \citet{davenport2016}. Similarly, producing an age for each star requires adopting a ``gyrochrone'' (gyrochronology isochrone) model prescription. Considerable effort has been made to explore the accuracy of such models for main sequence stars at a range of ages in the \Kepler era \citep[e.g.][]{mms+11,angus2015,douglas2016}. Most models produce similar age--rotation estimates for stars between roughly 500 Myr and a few Gyr. However, significant problems appear to exist for older, slower rotators, where a break in the angular momentum loss mechanism results in old stars (Solar age and older) with anomalously fast rotation periods \citep{van-saders2016}. As our sample of flare stars is biased towards the more active, younger main sequence stars, we do not expect this spin-down break to affect our results. We calculate ages using the \citet{mamajek2008} gyrochrone model. Future gyrochronology models will produce more accurate ages as a function of observed rotation periods, at which point we can update our FFD evolution relation accordingly. In the interim, this model will produce robust relative ages, and a reasonably accurate distribution of absolute ages as a function of the observed rotation period and stellar color.

To model the time dependent FFD we have adopted a power law decrease in the flare activity rate over time. Our chosen parameterization is supported by independent observations of the decline in other magnetic activity indicators over stellar age, including the fractional flare luminosity discussed in \S\ref{sec:fracL}. As a reminder, the FFD for a single star is defined  \citep[e.g. Eqn. 18 from][]{lme1976} as:
\begin{equation}
\log \nu = \alpha \, \log \varepsilon + \beta,
\end{equation}

\noindent
where $\nu$ is the cumulative rate of flares observed at a given energy, $\varepsilon$ is the flare event energy, and $\alpha$ and $\beta$ are the linear coefficients representing the power-law fit\footnote{Note: the use of coefficients $\alpha$ and $\beta$ are sometimes swapped in this equation.}.
To parameterize our time-evolving FFD as a function of stellar mass and flare energy, we have used the following model to fit the FFDs of all stars in our sample simultaneously:

\begin{eqnarray}
\label{eqn:model}
& \log \nu = a \log \varepsilon + b\\
{\rm where}& \nonumber \\
a =& a_1 \log t + a_2 m + a_3 \nonumber\\
b =& b_1 \log t + b_2 m + b_3 \nonumber 
\end{eqnarray}

\noindent
Note the first part of this equation again defines a the FFD power-law distribution for the reverse cumulative number of flares per day ($\nu$) as a function of flare event energy in erg ($\varepsilon$), written to take a linear form in log-log space with terms added to $a$ and $b$ that include a linear dependence on the $\log$ of the stellar age ($t$, in Myr), as well as the stellar mass ($m$, relative to solar). This equation allows us to fit the ensemble of individual stellar FFDs to determine $a_1, a_2, a_3, b_1, b_2, $ and $b_3$, and thus make predictions for the rate of flares as function of $\varepsilon, t, $ and $M_\odot$.

We explored other parameterizations of this FFD evolution model, including additional cross terms as a function of mass and age, e.g. $a_4 (m \times \log t)$. The Bayesian Information Criterion (BIC) did not yield a significant increase in the quality of the fit to the data given these extra degrees of freedom. Further, we currently have no theoretical basis for using a more complicated model for flare evolution as a function of time or mass. The possible break in the FFD power law for large energy flares, noted by \citet{davenport2016} and in our Figure \ref{fig:ffd1}, was not included in our parameterization of Equation \ref{eqn:model}. Instead we focus on reproducing the single dominant power law FFD relationship, which has been reported to characterize the distribution of flares on the Sun and solar-type stars over more than 10 orders of magnitude in event energy \citep[e.g. Fig. 9 from][]{shibayama2013}. We also did not attempt to include the ``saturated'' activity regime, suggested by \citet{davenport2016} for rapidly rotating stars with Rossby numbers lower than $\sim$0.03, in our FFD model. As \citet{davenport2016} notes, the break does not occur at the same Rossby number as other magnetic activity indicators (typically Ro=0.1) and the support for this break in the \Kepler flare data is currently tenuous. Exploring these additional model parameters is not technically difficult, but additional flare stars from missions like K2 and TESS are needed to determine if they are necessary.

Our model fitting in {\tt Python} took as inputs stellar ages in $\log(t/$Myr) and masses (in units of Solar mass), and all flare event energies in $\log(\varepsilon/$ erg), and produced the cumulative flare rate in $\log (\nu/day)$ over all flare energies measured for each star. 
Uncertainties in the flare rate were included in our fit, using the Poisson event counting uncertainty approximation from Equation 12 of \citet{gehrels1986}. No errors for the specific flare energies, or in our mass and age estimates were used in the fit. The model in Equation \ref{eqn:model} was fit to the entire flare star sample using an error weighted least-squares minimization to generate an initial guess for the six free parameters ($a_1, a_2, a_3,$ and $ b_1, b_2, b_3$).

\begin{table}
\caption{
Median parameters for the FFD evolution model described in Equation \ref{eqn:model} found from our MCMC exploration.
\label{tbl:params}
}
\footnotesize
\centering
\begin{tabular}{llll}
\tableline
$a_1=-0.07$  &  $a_2=0.79$ &    $a_3=-1.06$ \\
$b_1=2.01$ & $b_2=-25.15$ &  $b_3=33.99$  \\
\end{tabular}
\end{table}

\begin{figure*}[!t]
\centering
\includegraphics[width=6.25in]{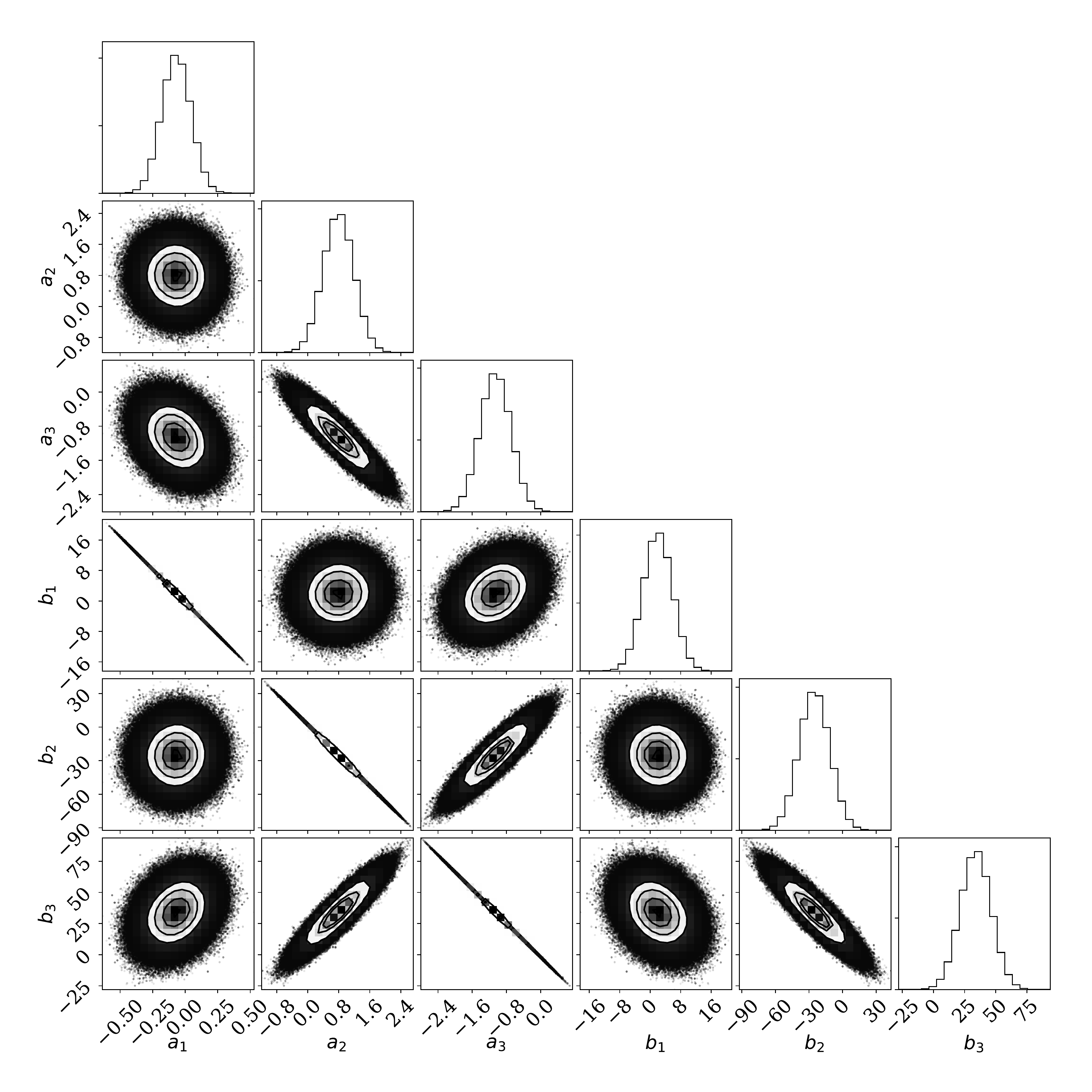}
\caption{
The standard {\tt emcee} sampler result {\tt corner} plot, showing the 1- and 2-dimensional posterior distributions for each free parameter in Eqn. \ref{eqn:model}. The density of points and contours correlate with the posterior probability distribution from a 100,000-step run of the {\tt emcee} sampler. Degeneracies are apparent between several parameters, seen here as very narrow distributions in three panels ($a_1, b_1$), ($a_2, b_2$), and ($a_3, b_3$), indicating our chosen model in Eqn. \ref{eqn:model} may have unnecessary complexity.
}
\label{fig:mcmc}
\end{figure*}

The initial fit for Equation \ref{eqn:model} was then refined using the Affine Invariant Markov Chain Monte Carlo (MCMC) ensemble sampler, {\tt emcee} \citep{emcee}. We first ran {\tt emcee}  for a ``burn-in'' phase of 500 steps using 100 walkers, seeded around the initial least-squares solution. The sampler was then run for 100,000 steps to explore parameter space. The MCMC chains were well converged for all 6 free parameters after the 100,000-step run, having an autocorrelation time of $\sim$60 steps for each parameter.
The final coefficients used for our FFD evolution model were determined using the median of the converged portion of the MCMC chains, and are given in Table \ref{tbl:params}. In Figure \ref{fig:mcmc} we also present the standard {\tt corner} plot \citep{corner}, which shows the 1- and 2-dimensional posterior probability density distributions for each free parameter for Equation \ref{eqn:model} from the MCMC sampler.

The 2-dimensional posterior distributions shown in Figure \ref{fig:mcmc} demonstrate that degeneracies are apparent between several parameters in our model. These can be identified as very narrow distributions in three panels of Figure \ref{fig:mcmc}: ($a_1, b_1$), ($a_2, b_2$), and ($a_3, b_3$). This can be interpreted as indicating that the model adopted in in Equation \ref{eqn:model} has unnecessary complexity, or that perhaps it could be re-cast into a more fundamental parameter space. As noted above, we have chosen the FFD evolution model in Equation \ref{eqn:model} for ease of implementation, and for lack of a better theoretical basis. The posterior distributions shown in Figure \ref{fig:mcmc} make it clear that other parameterizations should be explored in future studies.

\begin{figure*}[!t]
\centering
\includegraphics[width=2.3in]{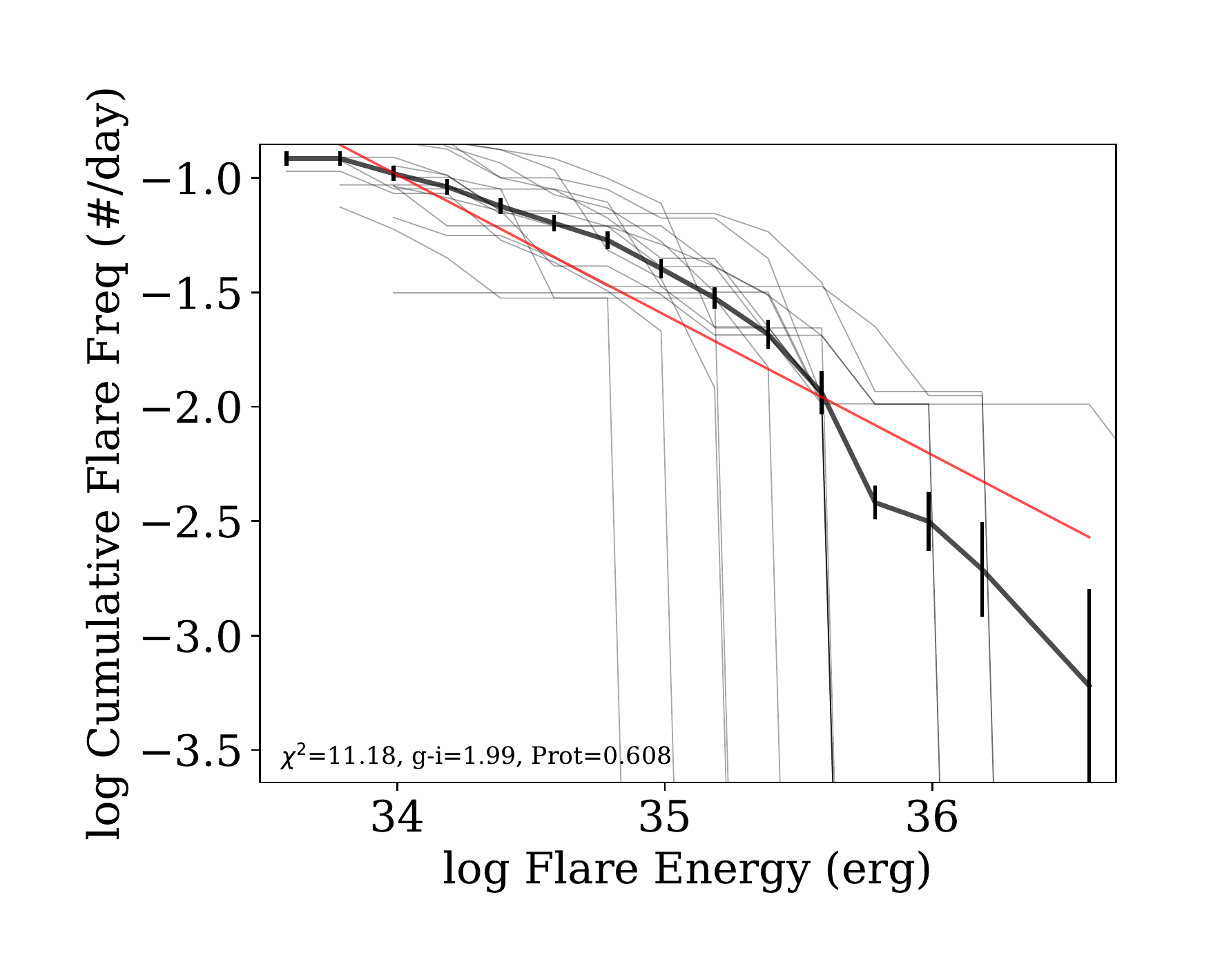}
\includegraphics[width=2.3in]{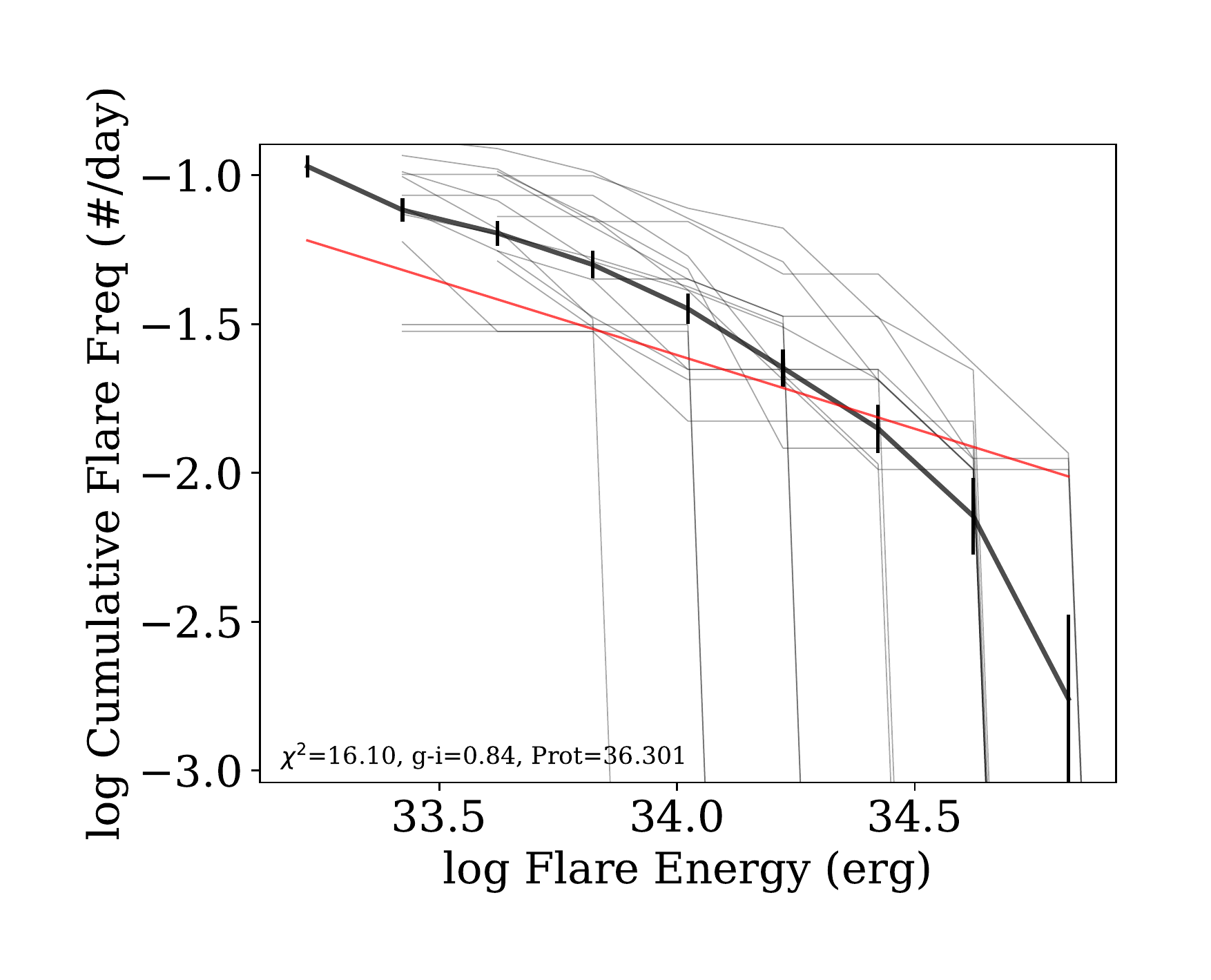}
\includegraphics[width=2.3in]{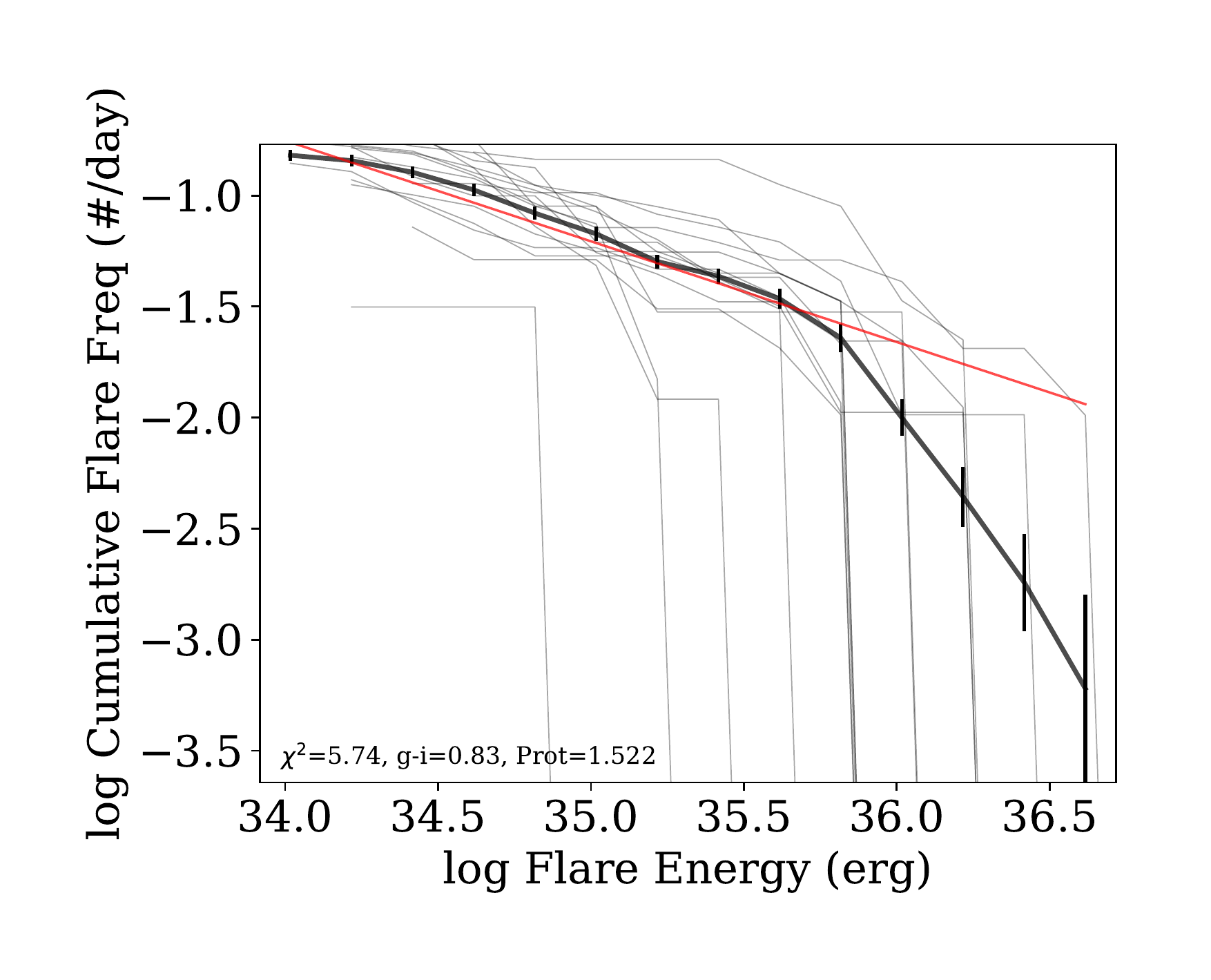}
\caption{
Flare frequency distributions as shown from Figure \ref{fig:ffd1} (black line), but with the final flare activity model from Equation \ref{eqn:model} evaluated for each star's mass and age (red line). Note this model was not fit for each star's FFD individually, but instead was fit to our entire sample of 347 stars.}
\label{fig:ffdmodel}
\end{figure*}

In Figure \ref{fig:ffdmodel} we demonstrate our FFD evolution model's ability to reproduce the observed FFDs for several stars. Here we show the combined FFD for the same three flare stars as in Figure \ref{fig:ffd1}, but with the predicted FFD from Equation \ref{eqn:model} overlaid. As discussed previously, we do not attempt to model any break in the FFD slope at high flare energies. The model reproduces the specific flare rates for each stars well, as well as the dominant slopes for the FFDs. These examples represent the ability of our flare evolution model to describe the FFD for a low-mass star at a given mass and age, and we anticipate this model will be helpful for future studies of the flare activity, such as exoplanet host evolution, or updating the predicted flare yields from surveys like LSST \citep{najita2016}.

We briefly note our lack of including null detections in this analysis. While our sample contains FFDs for 347 stars, there were nearly two orders of magnitude more stars with good colors (and therefore mass estimates) from \citet{davenport2016} and rotation periods from \citet{mcquillan2014} that did not meet our strict flare activity requirements. The flare injection tests from \citet{davenport2016} do also provide conservative estimates of the lowest energy flare that could have been detected in each object's \Kepler light curve with {\tt appaloosa}. We experimented with including these null detections in our model fitting as upper limits, using an observed flare rate of zero, and an uncertainty on the flare rate of 1 event per the total duration of observation ($\sim$4 years for most targets). However, we found two practical challenges that prevented us from using these targets in our model fitting: 1) it was not clear over what energy range in the FFD this upper limit should be considered for each object, and 2) these upper limits dominated the sample, and thus drove the model fitting to flatten the FFD evolution towards zero, systematically under-predicting the flare rates observed in our entire ``good'' sample of 347 stars. 
As such, while our model provides a useful parameterization of flare activity over time, it is only constrained for active stars and may over-estimate flare rates from inactive stars \citep[e.g.][]{hawley2014}.

Further, while our use of {\tt emcee} seeded with a least-squares fit follows standard approaches for fitting models to data in modern astronomy, it assumes Gaussian uncertainties to the data. The uncertainties in flare rates, however, are based on Poisson counting errors from \citet{gehrels1986}, and so Poisson regression approaches such as generalized linear modeling (GLM) may be better suited in. These approaches should give nearly identical fits for large samples, but will disagree for rare events (such as large-energy flares). Indeed, this may partially account for our over-estimation of flare rates for less active stars, as demonstrated in the next section.

\section{Exploring the Flare Rate Evolution}
\label{sec:model}

\begin{figure}[!t]
\centering
\includegraphics[width=3.5in]{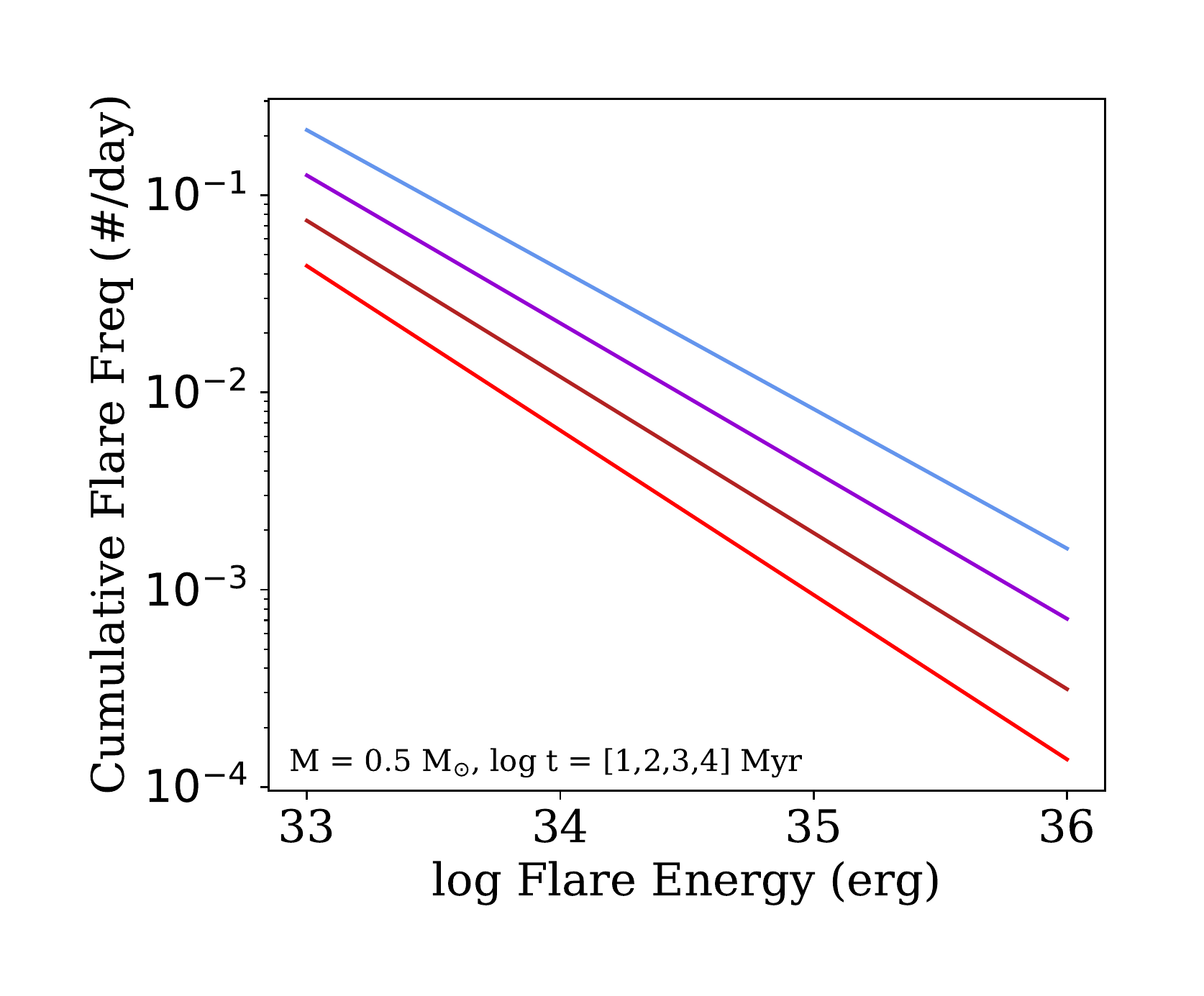}
\includegraphics[width=3.5in]{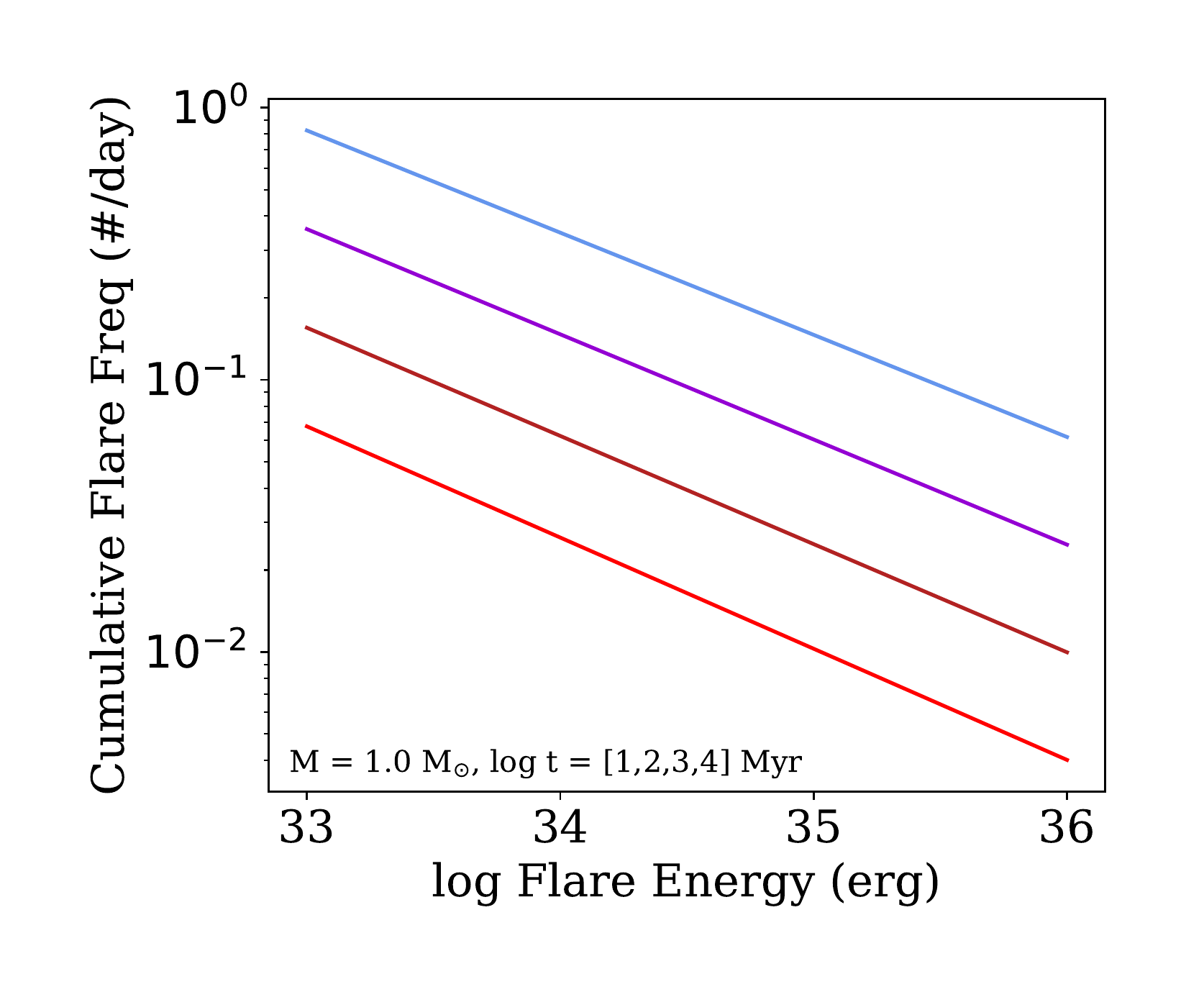}
\caption{
Our best-fit flare evolution model showing the predicted FFD evaluated at four age bins for a $0.5 M_\odot$ (top) and $1.0 M_\odot$ star (bottom). While the change in FFD slopes as a function of age is negligible, the difference as a function of mass is significant. Note: These FFDs are drawn at the same energy ranges for ease of comparison, but do not correspond to the specific flare energies detected at these masses.
}
\label{fig:model}
\end{figure}

Using our FFD evolution model generated from a sparse sample of field stars with varying masses and ages, we can now explore how a single star's flare activity evolves over its lifetime. In Figure \ref{fig:model} we show the FFD for a theoretical 0.5 M$_\odot$ star (approximately an M0 dwarf), and for a 1 M$_\odot$ star, both evaluated at four logarithmically spaced ages from 10 Myr to 1 Gyr. The solar-mass star's specific flare rate decreases more rapidly with time than the model M dwarf. The slope of the solar type star's FFD appears constant with time. In contrast, a very small change in the FFD slope is seen for the M dwarf, with the FFD getting steeper (fewer large flares) over time. However, given the small number of M dwarfs in our sample with old ages, the reliability of this change in slope is questionable. We find overall the FFD slope is effectively constant for stars at all ages.

Extending the FFD for a theoretical solar-mass star to the age of the Sun (here assumed to be 4.6 Gyr), we find our model over-predicts the rate of superflares determined for ``average Sun-like stars'' in \Kepler by \citet{shibayama2013}. Instead, our model closely fits their ``most active Sun-like'' sample. We believe this is due to our sample having insufficient numbers of slowly rotating G dwarfs to accurately predict the behavior of older stars. We also note a potential Malmquist bias in our flare rate determination exists due to the possible existence of activity cycles. As \citet{shibayama2013} and \citet{clarke2018} note, flare activity might vary by an order of magnitude over their activity cycles \citep[see also][]{veronig2002}, and so our sample may preferentially include stars near their highest activity states.

\begin{figure}[!t]
\centering
\includegraphics[width=3.5in]{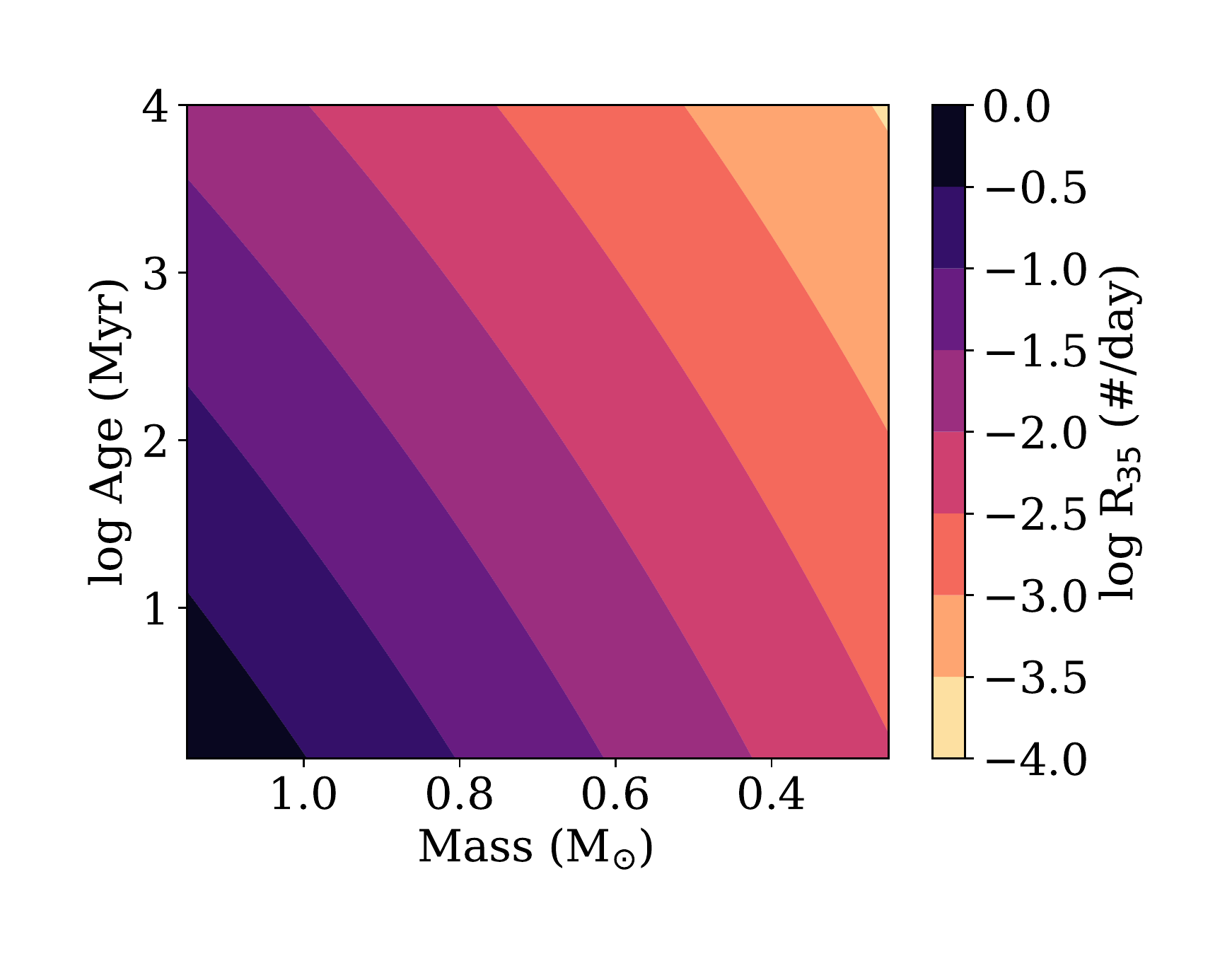}
\includegraphics[width=3.5in]{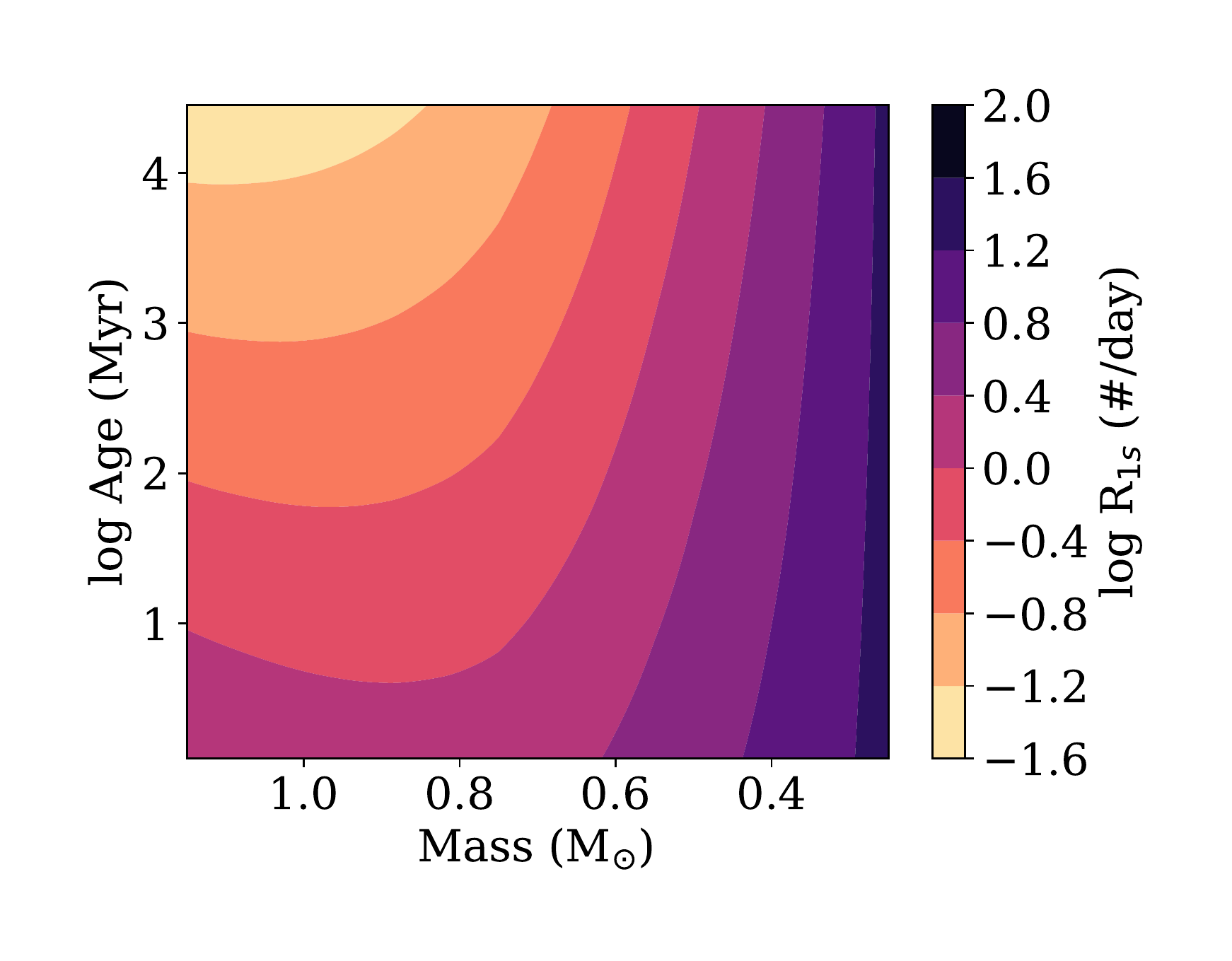}
\caption{
Our best-fit FFD evolution model evaluated over a grid of masses and ages, showing the cumulative flare rate $R_{35}$ for a fixed energy of $10^{35}$ erg (top), and at an equivalent duration of 1 second, i.e. $R_{1s} = $ 1 sec $\times L_{quies}$ (bottom). The latter clearly shows that low-mass stars produce more flares relative to their quiescent luminosity, while all stars show a decrease in their specific flare rate over time.
}
\label{fig:grid}
\end{figure}

Our FFD evolution model can also be used to predict {\it other} flare activity metrics described in this paper. For example, in Figure \ref{fig:grid} we display the mass + age dependence our models predicts for $R_{35}$, the specific flare rate at an absolute event energy of $\log \varepsilon = 35$ erg, and $R_{1s}$, the metric outlined in \S\ref{sec:rate} that scales with the quiescent luminosity of the star. The specific flare rate here is determined by evaluating the FFD at a given energy over a dense grid of ages and masses, allowing us to draw this phenomenological surface of flare activity as a function of mass and age. 
Encouragingly, the $R_{1s}$ surface shown in Figure \ref{fig:grid} demonstrates the expected behavior described in \S3.2, where low-mass star's flare activity declines more slowly with age.

Using our model to compute the fractional flare luminosity, $L_{fl}/L_{Kp}$, requires more care due to the considerations described in \S\ref{sec:fracL}. For example, the energy range of flare events that the model FFD is integrated over must be chosen at every age and mass. To match observations it is important to pick an energy range of flares that scales with each star's quiescent luminosity. The fractional flare luminosity was computed using a preliminary version of our FFD evolution model for \citet{clarke2018}, which examined the change in $L_{fl}/L_{Kp}$ between stars of varying mass ratios over a 1 Gyr timespan. This was computed at each age by integrating the FFD over six orders of magnitude in flare event energy, a larger dynamic range of events than typically observed for stars in \Kepler, even very active stars like GJ 1243 \citep{hawley2014}. However, the same range of energies was used for both stars, regardless of the mass ratio considered. \citet{clarke2018} found that the ratio of $L_{fl}/L_{Kp}$ between stars was roughly constant over time (up to 1 Gyr), which is again due to the model FFD slope remaining effectively the same at all ages for a given star.

Our results from Figures \ref{fig:model} and \ref{fig:grid} show that flare activity in Solar-mass stars decreases faster than for late-type stars (e.g. M dwarfs). As the surface rotation rate is directly connected to the strength of the stellar magnetic dynamo \citep{skumanich1972}, this finding is in broad agreement with the age--rotation--activity paradigm (gyrochronology) that shows faster spin-down for higher mass stars. 
Indeed, flares may play a very central role in the age--rotation--activity landscape. Large flares and their associated coronal mass ejections may be a key factor in driving stellar winds and thus stellar angular momentum loss \citep{johnstone2015b}. This may explain why, for example, flare activity does not appear to ``saturate'' at Ro$\sim$0.1 (Figure \ref{fig:rossby1}).

\section{Discussion and Summary}
\label{sec:discussion}

Here we have presented a thorough discussion of three methods for quantifying and comparing the flare activity of stars from optical light curves. Each method has advantages and disadvantages in terms of ease of implementation and computation, robustness to incomplete or low signal-to-noise data, and in the ability to compare between stars of varying masses and distances. The specific flare rate and the fractional flare luminosity are recommended for applications needing a simple, single metric, while modeling the entire flare frequency distribution (FFD) is preferred for accurately describing the details of a single, well studied flare star. 
All methods we studied, however, show definitively that flare activity decreases for all low-mass field stars as their rotation rates decrease (i.e. increasing age). This is qualitatively in agreement with recent studies of declining flare activity with age shown in young open clusters \citep[e.g.][]{ilin2018}.

In the development of the fractional flare luminosity metric, we have also introduced the ``$\Psi$ factor''. This quantity was described first by \citet{lurie2015} as a bolometric correction for the fractional flare luminosity observed in a given passband, and has been demonstrated here for the \Kepler band using an isochrone to estimate the bolometric luminosity. Since the \Kepler band cover a wide range of the optical regime, this correction is fairly insignificant for correcting flare activity in this sample. However, such an approach will be critical for comparing the flare activity between observations from various traditional optical bands (e.g. $U$-, $B$-, and $V$-bands).

In our ensemble modeling of the FFD, we have assumed a simple parameterization to continuously describe the evolution of flare activity as a function of both stellar mass and age (Eqn. \ref{eqn:model}). This model is a simple extension of the FFD power law commonly observed, with extra terms added for log(age) and mass. We have identified several interesting aspects of the FFD that have not been described by this model, including the potential break in the power law at high event energies, as well as regime of saturated activity for rapidly rotating (young) stars. As the sample of robust flare stars from \Kepler, K2, and soon TESS grows, the parameterization of the FFD evolution model should be revisited, with these features added a additional free parameters. The axes of this model should also be reexamined, such as parametrizing the evolution in terms of Rossby number rather than age.

We find the FFD slope does not significantly change for stars as a function of their age. This suggests our FFD evolution model could be simplified in the future by keeping the slope fixed. This result also has potentially deep meaning for the physics of flares, indicating for example the same rules of self-organized criticality are at work regardless of the star's total magnetic field strength. It is not clear from our work at this time, however, over what exact energy range this slope is applicable, nor how that energy range changes with stellar age. Unfortunately \Kepler's red wavelength coverage makes detecting very small energy flares difficult, and so it is not clear if this single power law slope extends to low enough energies to be an important component of support for the lower stellar atmosphere \citep[e.g. ``Ellerman Bombs'';][]{hansteen2017}. 
However, if the nano/micro-flare rate evolves coherently with the larger events and super-flare rates observed here (i.e. if the power law is preserved to very low energy events), this atmospheric support mechanism must change as the total flare rate decreases. Otherwise the lower stellar atmosphere support would decrease, and the stellar radius could change over time.

We have also introduced a new metric for quickly quantifying flare activity, $R_{1s}$. Since this metric estimates the specific flare rate scaled to the quiescent luminosity of the star, it is well suited for comparing flare activity between stars of varying masses. $R_{1s}$ is also useful for considering the impact of flares on exoplanet atmospheres and habitability, as it provides the rate of flares having approximately the same incident flux that a habitable zone exoplanet would experience.

Finally we note that all temporal evolution implied in this work relies on the assumption that gyrochronology accurately sorts stars as as a function of their age. This neglects, for example, any dynamical evolution in binaries that can affect the present day rotation \citep[e.g.][]{lurie2017,clarke2018}. We have also not utilized the stellar open clusters available in the \Kepler field, which are ideal benchmarks for stars at fixed ages. The sample of flare stars found in these clusters in \Kepler is too small to be used for a statistical analysis. However, with a number of nearby young clusters available in K2, future work should focus on comparing the flare activity between stars of known ages \citep[e.g.][]{ilin2018}.

\acknowledgments
We thank the anonymous referee and the AAS Statistics Consultant whose comments helped us clarify the impact and limitations of this work.

JRAD is supported by an NSF Astronomy and Astrophysics Postdoctoral Fellowship under award AST-1501418.

JRAD acknowledges support from the DIRAC Institute in the Department of Astronomy at the University of Washington. The DIRAC Institute is supported through generous gifts from the Charles and Lisa Simonyi Fund for Arts and Sciences, and the Washington Research Foundation

Kepler was competitively selected as the tenth Discovery mission. Funding for this mission is provided by NASA's Science Mission Directorate.

\software{Python, IPython \citep{ipython}, NumPy \citep{numpy}, Matplotlib \citep{matplotlib}, SciPy \citep{scipy}, Pandas \citep{pandas}, Astropy \citep{astropy}, emcee \citep{emcee}, corner.py \citep{corner}, PARSEC \citep{bressan2012}, appaloosa \citep{davenport2016}}


\end{document}